\documentclass[conference]{IEEEtran}
\IEEEoverridecommandlockouts

\usepackage{cite}
\usepackage{amsmath,amssymb,amsfonts}

\usepackage{amsthm}
\usepackage{algorithm}
\usepackage{algpseudocode}
\usepackage{graphicx}
\usepackage{textcomp}
\usepackage{xcolor}
\usepackage{booktabs}
\usepackage{multirow}
\usepackage{array}
\usepackage{tikz}
\usetikzlibrary{positioning,arrows.meta,calc,decorations.pathreplacing}
\usepackage{url}

\theoremstyle{plain}
\newtheorem{lemma}{Lemma}
\newtheorem{proposition}{Proposition}
\newtheorem{corollary}{Corollary}
\theoremstyle{definition}
\newtheorem{definition}{Definition}
\newtheorem{assumption}{Assumption}
\newtheorem{problem}{Problem}
\theoremstyle{remark}

\newcommand{\Enode}{E_{\mathrm{node}}}
\newcommand{\Ebuf}{E_{\mathrm{buf}}}
\newcommand{\Eps}{E_{\mathrm{ps}}}
\newcommand{\Scoped}[1]{\mathsf{S}_{#1}}
\newcommand{\Elru}[1]{\mathsf{L}_{#1}}
\newcommand{\IO}{\mathrm{IO}}
\newcommand{\Fconn}{F_{\mathrm{conn}}}
\newcommand{\Ewl}{\mathcal{E}}
\newcommand{\wclq}{w^{\mathrm{clq}}}
\newcommand{\warc}{w^{\mathrm{arc}}}
\newcommand{\Fclq}{F_{\wclq}}
\newcommand{\Str}{\Sigma_{\mathrm{tr}}}
\newcommand{\HP}{\textsc{HyperPage}}
\newcommand{\SET}{\textsc{Set}}
\newcommand{\CLQ}{\textsc{Clq}}
\newcommand{\PFG}{\textsc{P05-G}}
\newcommand{\SHPG}{\textsc{Shp-G}}
\newcommand{\FREQ}{\textsc{Freq}}
\newcommand{\HEAT}{\textsc{Heat}}
\newcommand{\Ehyb}[2]{\mathsf{H}_{#1,#2}}
\newcommand{\ARC}{\textsc{Arc}}
\newcommand{\SETG}{\textsc{Set-G}}
\newcommand{\CLQG}{\textsc{Clq-G}}
\newcommand{\ARCG}{\textsc{Arc-G}}
\newcommand{\ARCQ}{\textsc{Arc-Q}}
\newcommand{\SETQ}{\textsc{Set-Q}}
\newcommand{\STAR}{\textsc{Starling}}
\newcommand{\ARCC}{\textsc{Arc-C}}
\newcommand{\ARCCF}{\textsc{Arc-CF}}
\newcommand{\PFC}{\textsc{P05-C}}

\algrenewcommand\algorithmicrequire{\textbf{Input:}}
\algrenewcommand\algorithmicensure{\textbf{Output:}}

\begin{document}

\title{HyperPage: Learned Page Layouts for Disk-Resident Graph Vector Indexes}

\author{\IEEEauthorblockN{Xinyu Zhang\textsuperscript{1}, Jingze Li\textsuperscript{1}, Baihua Zheng\textsuperscript{2}, Weiwei Sun\textsuperscript{1}\thanks{Corresponding author: Weiwei Sun.}}
\IEEEauthorblockA{\textsuperscript{1}Fudan University, Shanghai, China\qquad\textsuperscript{2}Singapore Management University, Singapore\\
\{zhangxinyu, jingzeli20, wwsun\}@fudan.edu.cn, bhzheng@smu.edu.sg}}

\maketitle

\begin{abstract}
Disk-resident graph indexes such as DiskANN store each vector with its graph edges as one record on SSD pages. A search needs the page of every record it expands, so its page reads depend on the page layout, that is, on which records share a page. Search agents trained with reinforcement learning query one such index run after run, and every graph traversal can be logged with the records it expands. We therefore use the logs of a finished run to lay out the index for a declared memory budget, so that later runs read fewer pages. In contrast, existing layouts such as Starling and MARGO are computed from the graph and the vectors and use no logs. Learning a layout from logs is challenging, because the logs must stay valid under every layout, the saving depends on what the budget caches, and the best layout is NP-hard to find. Our key observation is that DiskANN-style search routes on in-memory codes and on the records it expands, never on their page-mates. As a result, the logged traversals price every layout by replay, and cached records add no page reads. Accordingly, we propose HyperPage, which first builds layouts for the declared budget on the logged traversals of the uncached records. It then selects one layout, together with the split of the budget between hot records and a page cache, by replaying validation steps. In addition, a Lagrangian lower bound shows how far a layout may be from the optimum. We learn from one Search-R1 agent run and test on a later run with a different reward, over 1M and 21M passages. Across seven cache budgets, HyperPage reads 5.2\% to 21.2\% fewer pages than MARGO, and no compared layout reads fewer. The advantage is largest when the cache is small. For example, at 1\% of the 1M index, MARGO needs nearly 3 times the budget to read as few pages. On the 21M index, HyperPage also cuts the search time per step by up to 10.4\%. The saving holds for a second agent model and a second corpus.
\end{abstract}

\begin{IEEEkeywords}
vector search, disk-resident index, physical design, page layout, workload logs, agent search
\end{IEEEkeywords}

\section{Introduction}\label{sec:intro}

Vector databases answer approximate nearest neighbor (ANN) queries over collections that exceed main memory. The dominant single-node design keeps product-quantization (PQ) codes~\cite{jegou2011pq} in DRAM and stores each vector with its adjacency list in a proximity graph as one record on SSD~\cite{subramanya2019diskann}. A page holds $B$ records, and beam search needs the page of every node it expands. Which records share a page changes no answer, but it changes the page reads, so the page layout is a physical-design decision of the vector database. It matters most when DRAM is small next to the index. In our record format, a billion records fill about 683\,GB of pages, so a workstation with 64\,GB of DRAM~\cite{subramanya2019diskann} caches under 5\% of the index after 32\,GB of PQ codes.

\begin{figure}[t]
\centering
\begin{tikzpicture}[x=1cm,y=1cm,font=\scriptsize,
    nd/.style={circle,draw=black!50,fill=white,inner sep=0pt,minimum size=1.6mm},
    ex/.style={circle,draw=black,fill=black,inner sep=0pt,minimum size=1.8mm},
    rec/.style={draw=black!40,fill=white,text=black!45,minimum width=0.27cm,minimum height=0.30cm,inner sep=0pt},
    hit/.style={rec,draw=black,fill=orange!30,text=black,font=\scriptsize\bfseries},
    pg/.style={draw=black!40,rounded corners=1pt,minimum width=0.89cm,minimum height=0.38cm,inner sep=0pt},
    pgr/.style={pg,draw=black,line width=1.1pt},
    arr/.style={-{Stealth[length=1.5mm]},semithick,draw=black!80},
    ttl/.style={font=\scriptsize\bfseries,anchor=west,inner sep=0pt}]
  \node[ttl] at (0,2.95) {Finished run (logged)};
  \foreach \n/\x/\y in {s/0.15/1.45, a/0.75/2.15, b/0.85/0.95, c/1.5/1.65, d/1.65/0.55, e/2.3/2.2, f/2.45/1.15,
                        u1/0.2/2.45, u2/2.95/1.95, u3/2.95/0.7, u4/1.15/0.2}
    \coordinate (\n) at (\x,\y);
  \foreach \u/\v in {s/a,s/b,a/c,b/c,b/d,c/e,c/f,d/f,e/u2,f/u2,f/u3,d/u4,s/u1,a/u1,d/u3,b/u4}
    \draw[black!28] (\u) -- (\v);
  \foreach \n in {s,u1,u2,u3,u4} \node[nd] at (\n) {};
  \foreach \n/\p in {a/above left,b/below left,c/above,d/below,e/above,f/below right} \node[ex,label={[inner sep=0.6pt,font=\scriptsize\bfseries]\p:\n}] at (\n) {};
  \draw[arr,black!60] (-0.3,1.45) node[left,inner sep=1pt,text=black]{$q$} -- (s);
  \node[align=left,anchor=north west,inner sep=0pt] at (0,-0.05) {searches of one rollout group\\expand records \textbf{a}--\textbf{f}};
  \draw[arr,thick] (3.3,0.85) -- node[above,align=center,inner sep=2pt]{learn a\\layout} node[below,align=center,inner sep=2pt]{for DRAM\\budget $D$} (4.38,0.85);
  \node[ttl] at (4.45,2.95) {Later run (new reward)};
  \newcommand{\pagerow}[3]{\foreach \p in {0,...,3} \node[pg] at (4.92+0.95*\p,#1) {};
    \foreach \p in {#3} \node[pgr] at (4.92+0.95*\p,#1) {};
    \foreach \r [count=\i from 0] in {#2} {
      \pgfmathtruncatemacro{\pp}{\i/3}\pgfmathtruncatemacro{\ss}{\i-3*\pp}
      \pgfmathsetmacro{\xx}{4.92+0.95*\pp+0.28*(\ss-1)}
      \ifnum\pdfstrcmp{\r}{f}>0 \node[rec] at (\xx,#1) {\r};\else\node[hit] at (\xx,#1) {\r};\fi}}
  \node[anchor=west,inner sep=0pt] at (4.45,2.5) {incumbent layout};
  \pagerow{2.1}{a,g,h,b,c,i,d,j,k,e,f,l}{0,1,2,3}
  \node[anchor=east,inner sep=0pt,font=\scriptsize\bfseries] at (8.35,1.67) {4 page reads};
  \node[anchor=west,inner sep=0pt] at (4.45,1.25) {learned layout};
  \pagerow{0.85}{a,b,c,d,e,g,f,h,i,j,k,l}{0,1,2}
  \node[anchor=east,inner sep=0pt,font=\scriptsize\bfseries] at (8.35,0.42) {3 page reads};
  \node[anchor=west,inner sep=0pt,text=black!60] at (4.45,0.05) {SSD pages of $B=3$ records};
\end{tikzpicture}
\caption{The task, on a constructed example with $B=3$. The logs of a finished run decide which records share an SSD page for a DRAM budget $D$. With a read-once buffer, a later group that expands a to f reads 4 pages under the incumbent layout and 3 under the learned one.}
\label{fig:task}
\end{figure}
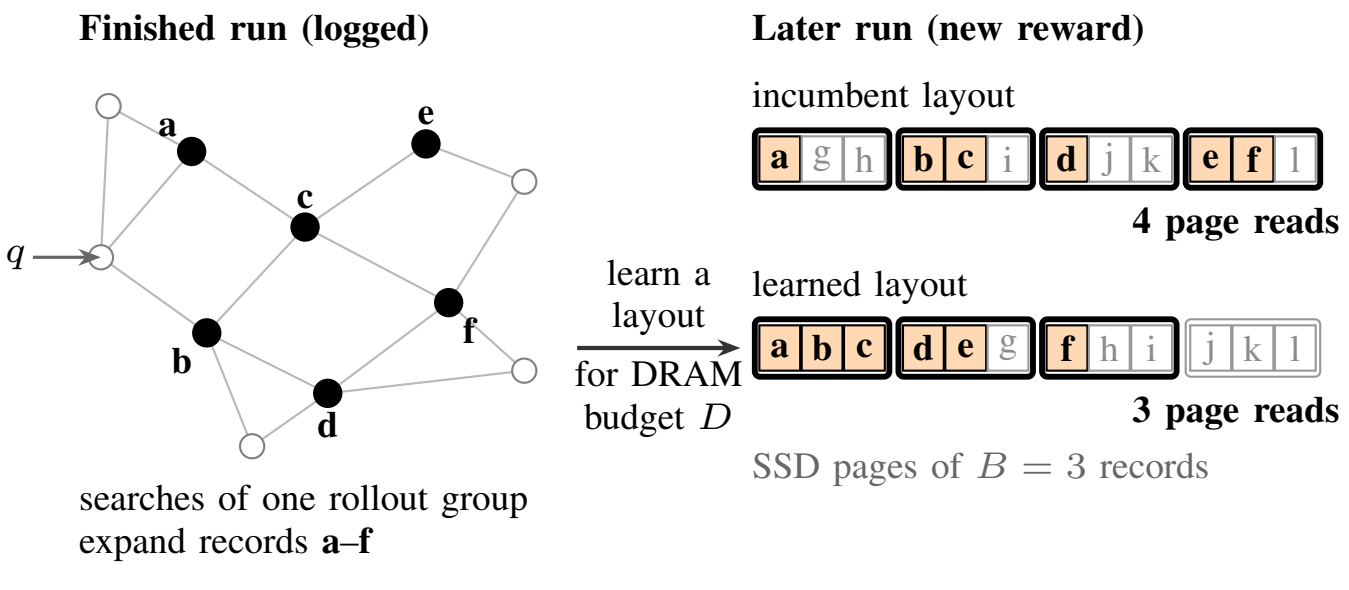

Search agents in training issue such queries at scale. An agent trained with group-relative policy optimization (GRPO)~\cite{shao2024deepseekmath,jin2025searchr1} runs for hundreds of steps, each rolling out 128 prompts eight times, and every search call can be logged with the graph nodes it expands. A later run searches the same corpus with the same tool under a new reward, and its traversals overlap heavily. On the run we test on, 83.9\% of the node expansions of a rollout group repeat a node that the group has already expanded.

We ask whether the logs of a finished run can lay out the index so that later runs read fewer pages under a declared DRAM budget (Fig.~\ref{fig:task}). A layout refined once then serves every later run of a retrieval tier shared by many runs and agents. A layout acts only on the accesses left uncached, so it stands in for budget. We study caches of 1\% to 25\% of indexes of 1M and 21M records. At equal shares, a larger index caches more of the workload: 1\% of the records serve 24.6\% of the test accesses at 1M and 36.4\% at 21M.

Existing layouts use no logs or logs of another workload. Starling and MARGO derive co-location scores from the graph and the vectors~\cite{wang2024starling,yue2025margo}, and MARGO's weights assume that queries follow the distribution of the indexed vectors. Workload hypergraph partitioning~\cite{eisenman2019bandana,kabiljo2017shp} treats logged requests as independent lookups. None of these works bounds the reads of the best layout from below, so none can say how much a better layout could still save.

Learning a layout from logs meets three obstacles. First, the logs must stay valid under every candidate layout, which page search breaks because its walks depend on the layout. Second, the saving depends on the buffer and the budget. Stock DiskANN reads once per expanded node, a cache of the whole index reads each page once under any layout, and caching hot records shrinks the gap between layouts. Third, the best layout is NP-hard to find once a page holds three or more records. Swap-based search can stop a factor of $B/2$ from the optimum, so a measured saving does not show how far a layout is from the best one.

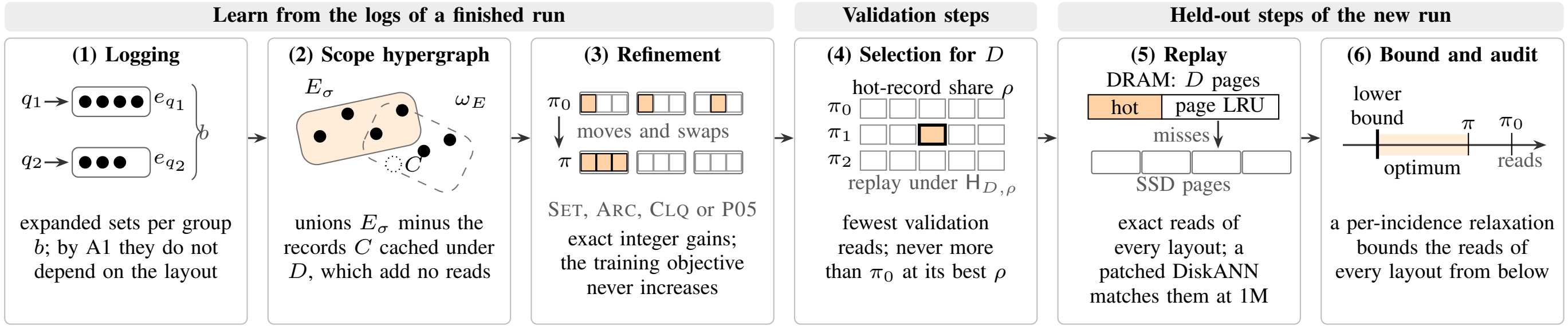
\begin{figure*}[t]
\centering
\begin{tikzpicture}[x=1cm,y=1cm,font=\scriptsize,
    dot/.style={circle,draw=black,fill=black,inner sep=0pt,minimum size=1.3mm},
    slot/.style={draw=black!40,fill=white,minimum width=0.17cm,minimum height=0.2cm,inner sep=0pt},
    hs/.style={slot,draw=black,fill=orange!35},
    pg/.style={draw=black!55,rounded corners=0.6pt,minimum width=0.56cm,minimum height=0.26cm,inner sep=0pt},
    cell/.style={draw=black!25,rounded corners=2pt,minimum width=2.8cm,minimum height=3.3cm,inner sep=0pt},
    phase/.style={fill=black!7,rounded corners=2pt,minimum height=0.34cm,inner sep=0pt,font=\scriptsize\bfseries},
    ttl/.style={font=\scriptsize\bfseries,anchor=north,inner sep=0pt},
    role/.style={align=center,text width=2.66cm,anchor=north,inner sep=0pt},
    tag/.style={align=center,text width=2.66cm,inner sep=0pt,text=black!65},
    arr/.style={-{Stealth[length=1.5mm]},semithick,draw=black!80}]
  \def\W{2.8}\def\G{0.24}
  \foreach \i in {0,...,5} {\coordinate (c\i) at ({\i*(\W+\G)+\W/2},0);}
  \node[phase,minimum width={(3*\W+2*\G)*1cm}] at ($(c1)+(0,3.7)$) {Learn from the logs of a finished run};
  \node[phase,minimum width={\W*1cm}] at ($(c3)+(0,3.7)$) {Validation steps};
  \node[phase,minimum width={(2*\W+\G)*1cm}] at ($(c4)!0.5!(c5)+(0,3.7)$) {Held-out steps of the new run};
  \foreach \i in {0,...,5} \node[cell,anchor=north] at ($(c\i)+(0,3.46)$) {};
  \foreach \i/\t in {0/{(1) Logging}, 1/{(2) Scope hypergraph}, 2/{(3) Refinement}, 3/{(4) Selection for $D$}, 4/{(5) Replay}, 5/{(6) Bound and audit}}
    \node[ttl] at ($(c\i)+(0,3.36)$) {\t};
  \foreach \a in {0,...,4} \draw[arr] ($(c\a)+(\W/2+0.01,2.3)$) -- ($(c\a)+(\W/2+\G-0.01,2.3)$);
  \begin{scope}[shift={($(c0)+(-1.22,1.6)$)}]
    \foreach \y/\q in {1.12/1,0.42/2} {
      \node[anchor=west,inner sep=0pt] at (0,\y) {$q_\q$};
      \draw[arr] (0.28,\y) -- (0.56,\y);
      \draw[black!60,rounded corners=2pt] (0.6,\y-0.17) rectangle (1.5,\y+0.17);
      \node[anchor=west,inner sep=0pt] at (1.53,\y) {$e_{q_\q}$};}
    \foreach \x in {0.75,0.95,1.15,1.35} \node[dot] at (\x,1.12) {};
    \foreach \x in {0.75,0.95,1.15} \node[dot] at (\x,0.42) {};
    \draw[decorate,decoration={brace,amplitude=3pt},black!70] (1.98,1.32) -- (1.98,0.22) node[midway,right=2pt,inner sep=0pt]{$b$};
  \end{scope}
  \node[role] at ($(c0)+(0,1.42)$) {expanded sets per group $b$; by A1 they do not depend on the layout};
  \begin{scope}[shift={($(c1)+(-1.1,1.6)$)}]
    \fill[orange!15,rounded corners=5pt,rotate around={12:(0.75,0.8)}] (0.05,0.5) rectangle (1.45,1.12);
    \draw[black!55,rounded corners=5pt,rotate around={12:(0.75,0.8)}] (0.05,0.5) rectangle (1.45,1.12);
    \draw[black!55,dashed,rounded corners=5pt,rotate around={-25:(1.35,0.65)}] (0.85,0.28) rectangle (2.05,0.98);
    \foreach \x/\y in {0.3/0.72,0.62/0.98,0.95/0.75,1.25/1.02,1.5/0.55,1.8/0.68} \node[dot] at (\x,\y) {};
    \node[circle,draw=black,densely dotted,fill=white,inner sep=0pt,minimum size=1.8mm] (cc) at (1.15,0.4) {};
    \node[inner sep=0.5pt,anchor=west] at (cc.east) {$C$};
    \node[inner sep=0.5pt] at (0.3,1.28) {$E_\sigma$};
    \node[inner sep=0.5pt] at (2.05,1.12) {$\omega_E$};
  \end{scope}
  \node[role] at ($(c1)+(0,1.42)$) {unions $E_\sigma$ minus the records $C$ cached under $D$, which add no reads};
  \begin{scope}[shift={($(c2)+(-0.98,1.6)$)}]
    \foreach \y/\lst in {1.1/{1,0,0,1,0,0,0,1,0}, 0.42/{1,1,1,0,0,0,0,0,0}} {
      \foreach \p in {0,1,2} \node[pg] at (0.42+0.66*\p,\y) {};
      \foreach \h [count=\k from 0] in \lst {
        \pgfmathtruncatemacro{\pp}{\k/3}\pgfmathtruncatemacro{\ss}{\k-3*\pp}
        \ifnum\h=1 \node[hs] at ({0.42+0.66*\pp+0.18*(\ss-1)},\y) {};\else\node[slot] at ({0.42+0.66*\pp+0.18*(\ss-1)},\y) {};\fi}}
    \node[anchor=east,inner sep=1pt] at (0.1,1.1) {$\pi_0$};
    \node[anchor=east,inner sep=1pt] at (0.1,0.42) {$\pi$};
    \draw[arr] (-0.1,0.95) -- (-0.1,0.58);
    \node[anchor=west,inner sep=0pt,text=black!65] at (0.1,0.76) {moves and swaps};
  \end{scope}
  \node[tag] at ($(c2)+(0,1.47)$) {\textsc{Set}, \textsc{Arc}, \textsc{Clq} or \textsc{P05}};
  \node[role] at ($(c2)+(0,1.24)$) {exact integer gains; the training objective never increases};
  \begin{scope}[shift={($(c3)+(-1.0,1.6)$)}]
    \foreach \r/\lab in {0/{\pi_0},1/{\pi_1},2/{\pi_2}} {
      \node[anchor=east,inner sep=1pt] at (0.32,1.04-0.3*\r) {$\lab$};
      \foreach \c in {0,...,4} \draw[black!45,fill=white] (0.36+0.34*\c,0.93-0.3*\r) rectangle (0.66+0.34*\c,1.15-0.3*\r);}
    \draw[black,line width=1pt,fill=orange!35] (0.36+0.34*2,0.93-0.3) rectangle (0.66+0.34*2,1.15-0.3);
    \node[inner sep=0pt] at (1.2,1.27) {hot-record share $\rho$};
    \node[inner sep=0pt,text=black!65] at (1.2,0.16) {replay under $\Ehyb{D}{\rho}$};
  \end{scope}
  \node[role] at ($(c3)+(0,1.42)$) {fewest validation reads; never more than $\pi_0$ at its best $\rho$};
  \begin{scope}[shift={($(c4)+(-1.15,1.6)$)}]
    \node[anchor=south,inner sep=1pt] at (1.2,1.2) {DRAM: $D$ pages};
    \draw[black,fill=orange!35] (0.1,0.92) rectangle (0.95,1.2);
    \draw[black,fill=white] (0.95,0.92) rectangle (2.3,1.2);
    \node[inner sep=0pt] at (0.525,1.06) {hot};
    \node[inner sep=0pt] at (1.625,1.06) {page LRU};
    \draw[arr] (1.6,0.9) -- (1.6,0.58);
    \node[anchor=east,inner sep=1pt,text=black!65] at (1.55,0.74) {misses};
    \foreach \p in {0,...,3} \node[pg] at (0.42+0.58*\p,0.42) {};
    \node[inner sep=0pt,text=black!65] at (1.2,0.15) {SSD pages};
  \end{scope}
  \node[role] at ($(c4)+(0,1.42)$) {exact reads of every layout; a patched DiskANN matches them at 1M};
  \begin{scope}[shift={($(c5)+(-1.25,1.6)$)}]
    \fill[orange!15] (0.5,0.5) rectangle (1.55,0.74);
    \draw[arr] (0.1,0.62) -- (2.45,0.62);
    \node[anchor=north east,inner sep=1pt,text=black!65] at (2.45,0.58) {reads};
    \draw[black,line width=1pt] (0.5,0.46) -- (0.5,0.8) node[above,inner sep=1pt,align=center]{lower\\bound};
    \draw[black] (1.55,0.5) -- (1.55,0.76) node[above,inner sep=1pt]{$\pi$};
    \draw[black] (2.05,0.5) -- (2.05,0.76) node[above,inner sep=1pt]{$\pi_0$};
    \node[inner sep=0pt,anchor=north] at (1.02,0.46) {optimum};
  \end{scope}
  \node[role] at ($(c5)+(0,1.42)$) {a per-incidence relaxation bounds the reads of every layout from below};
\end{tikzpicture}
\caption{Overview of \HP{}. Components (1) to (3) learn candidate layouts from the logs of a finished run. Component (4) selects a layout and a hot-record share on validation steps of the new run, and (5) and (6) price the deployment on held-out steps and bound the reads of every layout from below.}
\label{fig:overview}
\end{figure*}

Our key observation is that DiskANN-style search routes on in-memory codes and on the adjacency lists of the records it expands, never on their page-mates. The expanded sets logged in one run therefore price every layout exactly under read-once buffers, and under a shared cache in the logged order. This removes the first obstacle. It also shows that records held in a memory cache add no page reads, wherever they sit. For the second obstacle, \HP{} builds layouts for a declared budget on the logged unions of the uncached records. It then replays validation steps of the next run to select the layout and the split of the budget between hot records and a page LRU (Fig.~\ref{fig:overview}). For the third, a per-incidence Lagrangian relaxation bounds the reads of every layout from below, so a measured layout can be placed against the optimum.

We train on one run of a Search-R1 agent and test on 16,640 rollout groups of a later run with another reward, whose prompts never occur in training. \HP{} reads fewer pages than MARGO at every budget, 16.4\% to 21.2\% fewer at 1M and 5.2\% to 16.9\% fewer on the full corpus, and no compared layout reads fewer. At 1\% of either index, MARGO needs 1.5 to 3.0 times the budget to read as few pages. In a 12-thread DiskANN-based engine with equal measured memory, \HP{} cuts the median search time per step by up to 10.38\% on the full corpus. This engine with \HP{}'s layout is also faster than MARGO's page-search engine with either layout. The saving is largest at small cache shares. When the cache serves most accesses, a refinement that ignores the budget reads more pages than MARGO, so the budget must enter refinement. The saving holds for a second agent model and a second corpus.

Our contributions are as follows.
\begin{itemize}
\item \textbf{A replay model that prices every layout from one run's logs.} We state conditions under which the page reads of a buffer scope equal the connectivity of the union of its logged expanded sets, for every layout and arrival order. A patched DiskANN issues exactly the replayed reads.
\item \textbf{Cache-aware refinement and selection for a declared budget.} \HP{} builds layouts on the unions of the uncached records and selects the layout and cache split by replay on validation steps, with MARGO, Starling and MARGO's partitioner on the uncached records among the candidates. It reads fewer pages than MARGO at all seven budgets we test.
\item \textbf{A map of what can be guaranteed.} Under read-once scopes, every layout reads at most $B$ times the optimum, swap descent can stall at $B/2$, a matching reaches $\lceil B/2\rceil$, and the problem is APX-hard. A per-incidence Lagrangian bound places the 1M test optimum between 543.2 and 974.4 reads per group. Under a shared LRU no ratio in $B$ exists, and a window bound holds there.
\end{itemize}

\section{Related Work}\label{sec:related}

\paragraph*{Layouts and engines of disk-resident graph indexes}
Existing layouts of disk-resident graph indexes come from the graph and the vectors and target single queries. \HP{} instead learns a fixed graph's layout from a finished run's traversals and prices each candidate exactly under a declared buffer and budget. Starling shuffles records into blocks to maximize the in-block share of out-neighbors and proves the problem NP-hard with no finite-factor approximation~\cite{wang2024starling}. MARGO weights arcs by the monotonic search paths they support, fills pages greedily, and observes that more co-located edges need not mean fewer random I/Os~\cite{yue2025margo}. The SectorPack routine of OOD-DiskANN~\cite{jaiswal2022ooddiskann}, DiskANN++~\cite{ni2023diskannpp}, PageANN~\cite{kang2025pageann}, GPS~\cite{yin2026gps}, BAMG~\cite{li2026bamg} and VeloANN~\cite{zhao2026veloann} change the packing, graph, record format or edges, and DGAI~\cite{lou2025dgai} and GoVector~\cite{zhou2025govector} buffer the pages of a search. Only versions 1 and 2 of the LSM-VEC preprint use queries, reordering a dynamic disk graph under a query-sampled edge heatmap~\cite{zhong2025lsmvec} that we port as \HEAT{}.

\paragraph*{Batch and agent workloads}
Agent search systems change how a batch is searched or cached, not the page layout. CoWalk runs an agent batch as cooperating walks that read each page once and rank every vector of a fetched page~\cite{liang2026cowalk}, so its walks depend on the layout. DeepResearchGym serves deep-research agents from a DiskANN index~\cite{coelho2025deepresearchgym}, and InfoAgent and SemDN cache results or chunks~\cite{zhang2025infoagent,hua2026semdn}, which compose with a page layout.

\paragraph*{Workload hypergraphs and set objectives}
In workload-driven partitioning the application fixes each access set, while \HP{}'s sets are unions of graph traversals that stay valid under every layout only when walks do not depend on it. Connectivity partitioning of hypergraphs is classical~\cite{karypis1999hmetis,gottesburen2024mtkahypar}. Bandana minimizes the mean number of NVM blocks that each logged embedding lookup touches~\cite{eisenman2019bandana}, and MaxEmbed places co-appearing embeddings on SSD pages with replication~\cite{fan2024maxembed}. SHP minimizes a probabilistic fanout of logged queries~\cite{kabiljo2017shp}. Schism partitions a co-transaction graph~\cite{curino2010schism}, and Qd-tree~\cite{yang2020qdtree} and PAW~\cite{li2022paw} minimize the blocks a predicate workload accesses. We adopt SHP's objective at $p=0.5$ as \PFG{}.

\paragraph*{Trace-driven clustering for graph traversals}
\HP{} extends trace-driven clustering of graphs to ANN walks on a fixed graph, without replication. Object-graph clustering partitions graphs weighted by successive-access probabilities or trace-derived edges~\cite{tsangaris1991stochastic,tsangaris1992performance,gerlhof1993partition}, and Nodine et al.\ bound the speed-up of blocking for external graph search~\cite{nodine1996blocking}. ICBL and DiskGNN lay out graphs on disk~\cite{yasar2017icbl,liu2025diskgnn}, and SlashBurn, Gorder and near-neighbor reordering order vertices in memory~\cite{lim2014slashburn,wei2016gorder,coleman2022reordering}.

\paragraph*{Workload-driven adaptation of vector indexes}
Unlike systems that adapt a vector index to its workload, \HP{} changes only the node-to-page assignment of a fixed graph index. ConStore co-locates co-retrieved documents in content segments and rewrites them under drift~\cite{zhuang2026constore}. Other systems adapt partitions, trees or edges~\cite{mohoney2023hqi,mohoney2025quake,wei2025rotlex,zhang2026pigr,abuzakuk2026catapult}, storage requests~\cite{wan2026warp}, result caches~\cite{li2026anncache}, HNSW page placement~\cite{shim2025turbocharging}, tier placement~\cite{almamun2026coloc} or the graph itself~\cite{chen2024roargraph} to the workload. Replay prices hypothetical page layouts as what-if calls price index designs~\cite{chaudhuri1998whatif}, here exactly under Assumption~\ref{as:a1}.

\section{Preliminaries and Problem Definition}\label{sec:prelim}

We write $G$ for the index graph and $\mathcal H$ for the workload hypergraph.

\begin{definition}[Data, index, records, pages]\label{def:data}
The dataset is $X=\{\mathbf x_v\}_{v\in V}\subset\mathbb R^d$ with $n=|V|$. The proximity graph $G=(V,A)$ is fixed, with out-neighborhoods $N^+(v)=\{u:(v,u)\in A\}$ and $|N^+(v)|\le R$. Node $v$ is stored as one record of $s$ bytes that holds $\mathbf x_v$ and $N^+(v)$. With $c$ bytes per vector component and $L_2$ distance, $s=cd+4(R+1)$. A page has $P$ bytes and holds $B=\lfloor P/s\rfloor$ records.
\end{definition}

\begin{definition}[Layout, connectivity]\label{def:layout}
A layout is a map $\pi:V\to[k]$ with $k=\lceil n/B\rceil$ and $|\pi^{-1}(p)|\le B$ for every page $p$. $\Pi_B$ denotes the set of such layouts. For $S\subseteq V$, the page counts are $c_{S,p}(\pi)=|S\cap\pi^{-1}(p)|$, the page set is $\pi(S)=\{\pi(v):v\in S\}$, and the connectivity is $\lambda_\pi(S)=|\pi(S)|$.
\end{definition}

\begin{definition}[Logged runs, requests, groups]\label{def:logs}
A logged run has numbered steps. In a step, each prompt is rolled out $\nu$ times. A rollout proceeds in turns, and in a turn it may issue one tool call with a list of query strings. A \emph{request} is one tool call. A \emph{sub-query} $q$ is one element of its list. A \emph{group} $b$ is the set of sub-queries of one (step, prompt) over its $\nu$ rollouts. On our logs $\nu=8$, a step has 128 prompts, and a request holds 1.68 sub-queries on average.
\end{definition}

\begin{definition}[Beam search, expanded sets]\label{def:beam}
Sub-query $q\in\mathbb R^d$ (a string after embedding) runs best-first beam search from a fixed entry point with candidate-list size $L$ and beam width $W$, with routing distances from in-memory PQ codes. Round $t$ expands the beam $e^{(t)}_q$ with $|e^{(t)}_q|\le W$. The expanded set is $e_q=\bigcup_te^{(t)}_q$. Full-precision vectors from fetched records only re-rank the final candidates.
\end{definition}

We assume no node cache. A fixed cache $C$ replaces each $e_q$ by $e_q\setminus C$ and changes nothing else. Expanded nodes are marked visited, so the rounds are disjoint. Four assumptions describe the engines for which logged accesses remain valid under a new layout and buffers read each page once.

\begin{assumption}[A1: layout-independent walk]\label{as:a1}
For every sub-query $q$, every round $e^{(t)}_q$ is a function of $(q,G,L,W)$ and the PQ codes alone. It depends neither on $\pi$, nor on buffer contents, nor on other sub-queries. A1 holds when routing uses only the in-memory codes and the adjacency lists in the records of expanded nodes. Ties between equal PQ distances must break by original node identifiers, and co-resident records must never enter a candidate list.
\end{assumption}

\begin{assumption}[A2: reads of expanded nodes only]\label{as:a2}
The engine reads whole pages, only pages in $\pi(e_q)$, and holds page $\pi(v)$ when it expands $v\in e^{(t)}_q$. It does not prefetch.
\end{assumption}

\begin{assumption}[A3: cold start]\label{as:a3}
The buffer is empty when its sub-query or scope starts.
\end{assumption}

\begin{assumption}[A4: read-once scope buffer]\label{as:a4}
Within a scope, a page that has been fetched, or is in flight, serves every later request for that page by any sub-query of the scope until the scope ends. There is no eviction within the scope, and concurrent requests for an in-flight page are coalesced into one read.
\end{assumption}

\begin{definition}[Engines and buffer scopes]\label{def:engines}
$\Enode$ reads one page per expanded node, so $\IO_{\mathrm{node}}(\pi;q)=|e_q|$. $\Ebuf$ keeps every fetched page until the sub-query ends and coalesces reads in flight. A \emph{scope} $\sigma$ is a set of sub-queries, such as one request, one group $b$, one turn across a step, one step, or one session. The scoped buffer $\Scoped{\sigma}$ satisfies A4 for $\sigma$, and the sub-queries of $\sigma$ walk independently. $\Elru{D}$ is one LRU page cache of $D$ pages shared by all sub-queries in a given arrival order. $\Eps$ is page search. Every record of a fetched page is scored and may enter the candidate list, so $e_q=e_q(\pi)$ and A1 fails.
\end{definition}

\begin{definition}[Keyed buffer, proposed]\label{def:keyed}
The keyed buffer is $\Scoped{b}$ for groups: the server keeps one page buffer per key (step, prompt) for the generation phase of that step, which every request must carry.
\end{definition}

\begin{definition}[Memory budget, hybrid cache]\label{def:budget}
A deployment declares a budget of $D$ pages of DRAM for index data, a share $\eta=D/k$ of the index. PQ codes stay in DRAM in every compared setting and are not charged to $D$. For a hot-record share $\rho\in[0,1]$, the hybrid cache $\Ehyb{D}{\rho}$ spends $\lfloor\rho D\rfloor$ pages on a static cache of the $B\lfloor\rho D\rfloor$ records most frequent in training, whose accesses read nothing. Its other pages form one LRU page cache shared by all sub-queries in a given arrival order.
\end{definition}

$\Ehyb{D}{0}$ is $\Elru{D}$. At $\rho=1$ every uncached access reads its own page, so every layout reads the same. A pure page cache of $D\ge k$ pages reads each touched page once, and a split cache does so when the residual pages fit in its page part. We report $\eta$ of 1, 5 and 25\%, and on the full corpus also the 2\,GiB memory limit of MARGO's evaluation~\cite{yue2025margo}, which is 15\% of that index.

\begin{definition}[Scope hypergraph and objective]\label{def:objective}
A training workload is a set $\Str$ of scopes. Each scope $\sigma$ has the union $E_\sigma=\bigcup_{q\in\sigma}e_q$. Identical unions are merged, and $\omega_E>0$ is the normalized multiplicity of $E$, with $\sum_E\omega_E=1$. The scope hypergraph is $\mathcal H=(V,\Ewl,\omega)$ with $\Ewl=\{E_\sigma:\sigma\in\Str\}$. Single-sub-query scopes give the per-query hypergraph. We define
\begin{equation}\label{eq:fconn}
\Fconn(\pi)=\sum_{E\in\Ewl}\omega_E\bigl(\lambda_\pi(E)-1\bigr).
\end{equation}
\end{definition}

\begin{problem}[Learned page layout of a fixed disk-resident proximity graph]\label{prob:layout}
\emph{Input:} the fixed graph $G$ and record format, the page capacity $B$, and the weighted unions $\Ewl$ of a finished run.
\emph{Output:} a layout $\pi\in\Pi_B$.
\emph{Objective:} $\min_{\pi\in\Pi_B}\Fconn(\pi)$.
\emph{Constraints:} $\pi$ is a partition, so no record is replicated. The graph, record format, codes and search parameters stay unchanged.
\emph{Evaluation target:} the page reads of a later run, at stated buffer scopes and memory budgets. For a declared budget, Section~\ref{sec:budget} replaces $\Ewl$ by the unions of the uncached records.
\end{problem}

\section{Cost Model: What a Buffer Reads}\label{sec:cost}

The objective of Problem~\ref{prob:layout} is useful only if it equals what an engine pays. Every page-read number in this paper is attached to one engine through the next lemma, its corollary, or an explicit replay.

\begin{lemma}[Per-query cost model]\label{lem:cost}
Let A1, A2 and A3 hold for sub-query $q$. For every $\pi\in\Pi_B$, $\Ebuf$ reads $\IO_{\mathrm{buf}}(\pi;q)=\lambda_\pi(e_q)$ pages, and $\lceil|e_q|/B\rceil\le\lambda_\pi(e_q)\le|e_q|$. No engine that satisfies A1 to A3 reads fewer pages for $q$. The answers returned at fixed $(L,W)$ are the same under every layout.
\end{lemma}

\begin{proof}
By A1 the rounds are the same for every layout. By A2 and A3 every page of $\pi(e_q)$ is read at least once. $\Ebuf$ reads a page at its first reference and never again, so it reads $|\pi(e_q)|$ pages. A page holds at most $B$ nodes of $e_q$, which gives the lower bound. The candidate list depends only on PQ distances and $e_q$, and re-ranking uses the vectors of $e_q$, both fixed by A1.
\end{proof}

The same argument extends to scoped buffers once reads are coalesced within the scope.

\begin{corollary}[Scoped buffers]\label{cor:scope}
Let A1 and A2 hold for every sub-query of a scope $\sigma$, and let A3 and A4 hold for its buffer.
(a) $\Scoped{\sigma}$ reads exactly $\lambda_\pi(E_\sigma)$ pages for every $\pi$ and every arrival order of the sub-queries of $\sigma$. No engine that satisfies A1 to A3 for the scope reads fewer.
(b) $\lambda_\pi(E_\sigma)\le\sum_{q\in\sigma}\lambda_\pi(e_q)$.
(c) With an LRU buffer of $D\ge\lambda_\pi(E_\sigma)$ pages owned by $\sigma$ alone, in place of the unbounded buffer, the reads are still $\lambda_\pi(E_\sigma)$.
(d) Dropping a sub-query whose string repeats one of $\sigma$ changes neither $E_\sigma$ nor the reads.
\end{corollary}

\begin{proof}
(a) By A1 each $e_q$ is the same whether or not other sub-queries run, and in any order. By A4 each page of $\pi(E_\sigma)=\bigcup_q\pi(e_q)$ is read at most once, and by A2 and A3 at least once. (b) A union of page sets is no larger than the sum of their sizes. (c) At most $\lambda_\pi(E_\sigma)\le D$ distinct pages are ever referenced, so LRU never evicts. (d) By A1 equal strings have equal expanded sets, so the union is unchanged.
\end{proof}

Corollary~\ref{cor:scope} turns logged traversals into exact reads for every candidate layout, so layouts can be compared without running an engine. Part (a) needs no simultaneity, so a keyed buffer stays exact under any interleaving of a group's requests. Part (c) does not cover a cache shared by interleaved scopes. For $\Elru{D}$ no such identity holds, because its reads depend on the arrival order. Replay is still exact under A1 to A3 for any fixed order, and at $D\ge k$ nothing is evicted, so every layout reads each touched page once. Lemma~\ref{lem:cost} does not apply to $\Eps$, where the expanded sets depend on $\pi$. Engines that overlap expansion with outstanding reads, such as PipeANN~\cite{guo2025pipeann}, relax the best-first order of expansions, and A1 does not hold for them either. Thus Lemma~\ref{lem:cost} covers stock DiskANN as $\Enode$ and DGAI's buffer~\cite{lou2025dgai} as $\Ebuf$, and Corollary~\ref{cor:scope} covers the proposed keyed buffer and the batch cache of CoWalk~\cite{liang2026cowalk} on one thread. The page search of Starling, MARGO and CoWalk's cooperative walks~\cite{wang2024starling,yue2025margo,liang2026cowalk} is $\Eps$.

\paragraph*{When a layout can matter}
In the DiskANN main branch of May 2026, every frontier node receives its own read request, and on 10K test queries each of SIFT1M, DEEP1M and T2I-1M issued reads equal expanded nodes. A buffer such as $\Ebuf$ is therefore the precondition for any layout effect, including the 9\% that OctopusANN reports from page shuffle~\cite{li2026octopusann}. A layout also needs more than one record per page: a 384-dimensional int8 record takes 644 bytes, which gives $B=6$ at 4\,KB and $B=12$ at 8\,KB, the format we study.

\section{Analysis: What Reads Count}\label{sec:objectives}

By Corollary~\ref{cor:scope}, reads equal connectivity, so minimizing $\Fconn$ is the exact training cost under a scoped buffer. MARGO and Starling maximize weighted counts of co-located pairs. A pairwise objective can also be built from the logged sets themselves, which separates the information used from the form of the objective.

\begin{definition}[Pairwise co-location objectives]\label{def:pairwise}
For symmetric weights $w:\binom V2\to\mathbb R_{\ge0}$, let $F_w(\pi)=\sum_{\{u,v\}}w_{uv}\mathbf{1}[\pi(u)=\pi(v)]$, to be maximized. The clique expansion of the logged sets has $\wclq_{uv}=\sum_{E\ni u,v}\omega_E$ (\CLQ{}). Its arc-restricted version is $\warc_{uv}=\wclq_{uv}\cdot\mathbf{1}[(u,v)\in A\vee(v,u)\in A]$ (\ARC{}). Starling's objective uses the arc indicator as $w$, and MARGO's uses $w(u,v)=m(u,v)\,m(u)$ over arcs, where $m$ counts monotonic search paths~\cite{yue2025margo}. Minimizing $\Fconn$ is the set objective (\SET{}). For a set $E$, let $\phi_E(\pi)=\sum_p\binom{c_{E,p}(\pi)}2$, so that $\Fclq(\pi)=\sum_E\omega_E\phi_E(\pi)$.
\end{definition}

The set objective credits a page linearly in the members it holds, and the clique objective credits it quadratically. The following identity makes the difference exact.

\begin{proposition}[Set credit plus concentration bonus]\label{prop:bonus}
For every layout $\pi$ and every weighted family of sets, let $g(\pi)=\sum_E\omega_E(|E|-\lambda_\pi(E))$. Then
\begin{gather}
\Fclq(\pi)=g(\pi)+\Omega(\pi),\label{eq:bonus}\\
\Omega(\pi)=\sum_E\omega_E\sum_{p\in\pi(E)}\binom{c_{E,p}(\pi)-1}{2}\ge0,\notag
\end{gather}
and $\Omega(\pi)=0$ if and only if no page holds three or more members of one set with $\omega_E>0$.
\end{proposition}

\begin{proof}
Apply Pascal's rule $\binom c2=(c-1)+\binom{c-1}2$ to $c=c_{E,p}$ and sum over $p\in\pi(E)$.
\end{proof}

The credit $g$ counts the reads a layout saves against one page per member. Clique weight is also the first layout-dependent term of SHP's probabilistic fanout at small retention $\theta$~\cite{kabiljo2017shp}. Its case of two records per page shows where the form of the objective cannot matter.

\begin{corollary}[Two records per page]\label{cor:b2}
If $B=2$, then $\Fconn(\pi)=\sum_E\omega_E(|E|-1)-\Fclq(\pi)$ for every $\pi\in\Pi_2$. A layout minimizes $\Fconn$ if and only if its pages form a maximum-weight matching of the clique-expansion graph, completed by pairing unmatched nodes, which takes polynomial time~\cite{edmonds1965matching}.
\end{corollary}

\begin{proof}
At $B=2$ every $c_{E,p}\le2$, so $\Omega\equiv0$ and $\Fconn=\sum_E\omega_E(|E|-1)-g=\sum_E\omega_E(|E|-1)-\Fclq$. The co-located pairs of a layout in $\Pi_2$ form a matching of weight $\Fclq(\pi)$, and any matching extends to a layout by pairing unmatched nodes, which cannot decrease the nonnegative weight.
\end{proof}

Corollary~\ref{cor:b2} gives an exact optimizer at $B=2$, against which we calibrate the lower bound of this section.

Beyond two records per page, the bonus is limited by the members of one set that fit on a page, which the next proposition uses.

\begin{proposition}[Size and capacity bound]\label{prop:ratio}
Let $\bar b=\max_E\min(|E|,B)$ over the sets with $\omega_E>0$. For every $\pi\in\Pi_B$, $g(\pi)\le\Fclq(\pi)\le\frac{\bar b}2g(\pi)$. If $\pi_c\in\arg\max\Fclq$ and $\pi^\star\in\arg\min\Fconn$ over $\Pi_B$, then $g(\pi_c)\ge\frac2{\bar b}g(\pi^\star)$ and $\Fconn(\pi_c)-\Fconn(\pi^\star)\le(1-\frac2{\bar b})g(\pi^\star)$.
\end{proposition}

\begin{proof}
For $1\le c\le\bar b$, $c-1\le\binom c2\le\frac{\bar b}2(c-1)$. Summing gives the two-sided bound, and $g(\pi_c)\ge\frac2{\bar b}\Fclq(\pi_c)\ge\frac2{\bar b}\Fclq(\pi^\star)\ge\frac2{\bar b}g(\pi^\star)$.
\end{proof}

The clique objective is thus a $2/\bar b$ relaxation of the page savings of exact maximizers, with $\bar b=B$ on sets larger than a page. It says nothing about the heuristic arms, MARGO's weights or held-out scopes. The next result shows that the factor cannot be raised.

\begin{proposition}[Tightness]\label{prop:tight}
For every $B\ge3$ and $\varepsilon>0$ there is an instance on which $\Fclq$ has a unique maximizer $\pi_c$, up to renaming pages, and $\frac2B\le g(\pi_c)/g(\pi^\star)\le(1+\varepsilon)\frac2B$, where $\pi^\star$ minimizes $\Fconn$.
\end{proposition}

\begin{proof}[Proof sketch]
On $V=[B]\times[B]$, take the rows as sets of weight $1+\varepsilon$ and every pair of cells in one column as a set of weight 1. The row layout uniquely maximizes $\Fclq$ with $g=(1+\varepsilon)B(B-1)$, and the column layout has $g\ge B\binom B2$.
\end{proof}

The scope is a second axis. A per-query objective, set or pairwise, scores each sub-query of a scope on its own.

\begin{proposition}[Cross-query reuse]\label{prop:reuse}
For a scope $b$ and a layout $\pi$, let $r_{b,p}(\pi)=|\{q\in b:p\in\pi(e_q)\}|$. Then
\begin{equation}\label{eq:reuse}
\sum_{q\in b}\lambda_\pi(e_q)=\lambda_\pi(E_b)+\sum_{p\in\pi(E_b)}\bigl(r_{b,p}(\pi)-1\bigr).
\end{equation}
There exist workloads with $B=2$ for which every layout that minimizes the per-query objective $\sum_b\sum_{q\in b}\lambda_\pi(e_q)$ reads strictly more scoped pages $\sum_b\lambda_\pi(E_b)$ than some other layout.
\end{proposition}

\begin{proof}
Each page of $\pi(E_b)$ is counted once on the right of~\eqref{eq:reuse} by $\lambda_\pi(E_b)$ and $r_{b,p}-1$ more times by the sum, and once on the left for every sub-query of $b$ that touches it. For the inversion, take $B=2$, a scope with the set $\{v,z\}$, and a scope with the sets $\{u,v\}$ and $\{w,z\}$. The pages $\{u,v\},\{w,z\}$ cost 4 per query and 4 scoped, the unique per-query optimum, while $\{u,w\},\{v,z\}$ cost 5 and 3.
\end{proof}

The reuse term of~\eqref{eq:reuse} is what a scoped buffer saves, and it depends on the layout. Proposition~\ref{prop:reuse} is why \HP{} trains on scope unions: an objective that scores the sub-queries of a scope separately can miss that reuse.

Exact optimization is out of reach beyond two records per page, which is why we refine heuristically and audit the result with a lower bound.

\begin{proposition}[Complexity]\label{prop:complexity}
Over arbitrary weighted set families, minimizing $\Fconn$ on $\Pi_B$ is NP-hard for every fixed $B\ge3$, and it takes polynomial time for $B=2$.
\end{proposition}

\begin{proof}
The case $B=2$ is Corollary~\ref{cor:b2}. For $B\ge3$, take a graph $H=(U,E_H)$ with $|U|=Bk$ and use its edges as sets. Every page then holds $B$ nodes, and $\Fconn$ counts cut edges, up to the normalization of $\omega$. The cut is at least $|E_H|-k\binom B2$, with equality if and only if every page induces $K_B$. Deciding whether such a $K_B$-factor exists is NP-complete~\cite{kirkpatrick1978completeness}.
\end{proof}

Approximation is limited as well. For every fixed $B\ge3$, minimizing the reads of a read-once workload is APX-hard even on weighted pairs, so no polynomial approximation scheme exists unless P${}={}$NP. The reduction is from maximum independent set on 3-regular graphs~\cite{chlebik2006bounded,alimonti2000cubic}.

\begin{table}[t]
\centering
\caption{Worst-case reads against the optimum $R^\star$ under read-once scopes.}
\label{tab:bounds}
\footnotesize
\setlength{\tabcolsep}{3pt}
\begin{tabular}{@{}ll@{}}
\toprule
Layout & Reads at most \\
\midrule
Any layout (tight), so also Algorithm~\ref{alg:hyperpage} & $(U/L_{\mathrm{cap}})R^\star\le BR^\star$ \\
Some swap-local optima of $R$, every $B$ & $\frac B2R^\star$ reached \\
Maximum-weight matching, polynomial & $\lceil B/2\rceil R^\star$ \\
Exact maximizer of $\Fclq$ (from Proposition~\ref{prop:ratio}) & $\lfloor(\bar b+1)^2/4\rfloor R^\star/\bar b$ \\
Exact minimizer of the fanout at $\theta=\frac12$ & $(2-2^{1-B})R^\star$ \\
No polynomial algorithm, unless P${}={}$NP & $(1+\varepsilon_B)R^\star$ \\
\bottomrule
\end{tabular}
\end{table}

\paragraph*{What can be guaranteed}
Write $R(\pi)=\sum_E\omega_E\lambda_\pi(E)$ for the reads of a read-once workload and $R^\star$ for their minimum over $\Pi_B$. Page capacity alone bounds every layout.

\begin{proposition}[Capacity guarantee]\label{prop:cap}
Let $L_{\mathrm{cap}}=\sum_E\omega_E\lceil|E|/B\rceil$ and $U=\sum_E\omega_E\min(|E|,k)$. Every $\pi\in\Pi_B$ satisfies $L_{\mathrm{cap}}\le R^\star\le R(\pi)\le U\le B\,L_{\mathrm{cap}}$, and the factor $B$ is attained.
\end{proposition}

\begin{proof}
Per set, $\lceil|E|/B\rceil\le\lambda_\pi(E)\le\min(|E|,k)\le B\lceil|E|/B\rceil$. On $B$ disjoint sets of $B$ records, one set per page reads $B$, and one record of each set per page reads $B^2$.
\end{proof}

Table~\ref{tab:bounds} shows where \HP{} stands. Its refinement is swap descent, so it inherits only the first row. Swaps can stall at a factor of $B/2$. With $\lfloor B/2\rfloor$ disjoint sets of $2B$ records on $B$ pages, two records of every set on every page read $B/2$ times the optimum, and no exchange lowers the reads. Polynomial time reaches $\lceil B/2\rceil$: a maximum-weight matching of the clique expansion~\cite{edmonds1965matching}, packed $\lfloor B/2\rfloor$ pairs to a page, outweighs a random matching inside the pages of an optimal layout. The clique expansion of our unions does not fit in memory, so the matching stays a reference point. The surrogates differ sharply. The fanout credits a page between $\frac12$ and $1-2^{-B}$ per touched set, so its exact minimizer stays within a factor of two at every $B$, while the factor of the clique objective grows with $B$. This agrees with \textsc{P05} reading less than \CLQ{} at both scales in our experiments. For fixed $B\ge3$ the best polynomial factor therefore lies between $1+\varepsilon_B$, a small constant, and $\lceil B/2\rceil$, and this range is open. For a measured layout we instead use the ratio to a lower bound: any $L\le R^\star$ gives $L\le R^\star\le R(\pi)=\varrho(\pi)L$, so $\pi$ is within $\varrho(\pi)$ of the optimum. Next is a bound $L$ stronger than $L_{\mathrm{cap}}$ on our workloads. Let $\beta=B-1$, $h(t)=1/(1+t)$, $m_E=\min(\beta,|E|-1)$, and $\psi_E(g)=\min_{t\in\{0,\dots,m_E\}}h(t)+gt$.

\begin{proposition}[A lower bound valid for every layout]\label{prop:lb}
Partition the records into classes $C_1,\dots,C_r$ and write $c(u)$ for the class of $u$. For real $\gamma_{vE}$, one per record $v$ and set $E\ni v$, and for $\mu\in\mathbb R^r_{\ge0}$, let $w^\gamma_{vu}=\sum_{E\ni u,v}\omega_E\gamma_{vE}$. Then for every $\pi\in\Pi_B$ the reads $\Fconn(\pi)+\sum_E\omega_E$ satisfy
\begin{multline*}
\sum_E\omega_E\lambda_\pi(E)\ge\sum_v\Bigl[\sum_{E\ni v}\omega_E\psi_E(\gamma_{vE})\\
-\max_{\substack{T\subseteq V\setminus\{v\}\\|T|\le\beta}}\sum_{u\in T}\bigl(w^\gamma_{vu}-\mu_{c(u)}\bigr)\Bigr]-\beta\sum_i\mu_i|C_i|.
\end{multline*}
\end{proposition}

\begin{proof}
Let $S_v$ be the page-mates of $v$ and $t_E=|S_v\cap E|$, so $c_{E,\pi(v)}=1+t_E$ and $\sum_E\omega_E\lambda_\pi(E)=\sum_vr_v$ for $r_v=\sum_{E\ni v}\omega_Eh(t_E)$. As $t_E\le m_E$, $h(t_E)\ge\psi_E(\gamma_{vE})-\gamma_{vE}t_E$, and $\sum_{E\ni v}\omega_E\gamma_{vE}t_E=\sum_{u\in S_v}w^\gamma_{vu}$. Hence $r_v\ge\sum_{E\ni v}\omega_E\psi_E(\gamma_{vE})-\sum_{u\in S_v}(w^\gamma_{vu}-\mu_{c(u)})-\sum_{u\in S_v}\mu_{c(u)}$, and as $|S_v|\le\beta$ the middle sum is at most the maximum over $T$. Page-mates are mutual, so $\sum_v\sum_{u\in S_v}\mu_{c(u)}=\sum_i\mu_i\sum_{u\in C_i}|S_u|\le\beta\sum_i\mu_i|C_i|$.
\end{proof}

As in Lagrangian relaxation~\cite{fisher1981lagrangian}, each $\gamma_{vE}$ prices one page-mate of $v$ inside $E$, so the bound sees how a record's page-mates spread over its sets. The prices $\mu$ charge each class for its page-mate capacity $\beta|C_i|$, a constraint aggregation~\cite{elhallaoui2005dynamic}, and the inner maximum needs only each record's $\beta$ heaviest partners per class. We use degree classes and projected subgradient steps. The worst-case gap grows with $B$, because the proof also holds for fractional page-mates that need not be mutual. On the vertex stars of a $\delta$-regular graph of girth above $B$ with $\delta\ge(B+1)/2$~\cite{erdos1963regulare}, whose edges are the records, spreading each record evenly over its $2\delta-2$ neighbours costs at most $4(B+1)/(B(B+2))$ per record, while a page of $j$ records touches at least $j+1$ sets. The optimum thus exceeds every bound of Proposition~\ref{prop:lb} by a factor of at least $(B+2)/4$.

\paragraph*{Shared caches}
None of this transfers to a page LRU, whose misses depend on the order of accesses. With $B=2$ and one LRU page, the trace $(a,b)^T$ misses once on the pages $\{a,b\},\{c,d\}$ and $2T$ times on $\{a,c\},\{b,d\}$. No function of $B$ therefore bounds the ratio of two layouts, and the sum of cold scope bounds does not bound the misses either. A weaker bound survives.

\begin{proposition}[Window bound]\label{prop:window}
Fix a record cache and a page cache of $H$ pages, empty at the start. Cut the access trace into consecutive windows with uncached record sets $S_1,\dots,S_J$. Every layout misses at least $\lceil|S_1|/B\rceil+\sum_{j\ge2}\max(0,\lceil|S_j|/B\rceil-H)$ times.
\end{proposition}

\begin{proof}
Window $j$ touches at least $\lceil|S_j|/B\rceil$ pages, and at most $H$ of them are resident when it starts.
\end{proof}

\section{HyperPage}\label{sec:method}

Fig.~\ref{fig:overview} shows the six components of \HP{}. Section~\ref{sec:logging} covers (1) and (2), Sections~\ref{sec:gains} and~\ref{sec:partition} cover (3), and Section~\ref{sec:budget} adds the budget and covers (4). Section~\ref{sec:engine} covers (5), and Proposition~\ref{prop:lb} gives the bound of (6).

\begin{algorithm}[t]
\caption{Refinement$(F,\mathcal H,\omega,B,\pi_0,\tau,\epsilon)$}\label{alg:hyperpage}
\begin{algorithmic}[1]
\Require objective $F$ to minimize, weighted unions $\Ewl$, capacity $B$, incumbent $\pi_0$, passes $\tau$, tolerance $\epsilon$
\Ensure layout $\pi\in\Pi_B$
\State $\pi\gets\pi_0$; $O\gets$ random order of the covered nodes
\For{$i=1,\dots,\tau$}
  \ForAll{$v\in O$}
    \State $\mathcal P\gets$ the $\bar p$ candidate pages of $v$
    \State among moves into free slots of $\mathcal P$ and swaps with occupants of $\mathcal P$, take the most negative exact change $\Delta F$
    \State apply it if $\Delta F<0$
  \EndFor
  \State \textbf{break} if the pass gain is at most $\epsilon F(\pi)$
\EndFor
\State \Return $\pi$
\end{algorithmic}
\end{algorithm}

\subsection{Logging and the Scope Hypergraph}\label{sec:logging}
For each training sub-query, the logger records the expanded set $e_q$ in expansion order and the group that issued it. By A1 the expansion depends only on the codes and the graph, so the logs are valid for every layout. The optimizer receives the unions $E_\sigma$ of the training scopes, merged with multiplicity. A group union summarizes a log and assumes nothing about how requests arrive, so a layout trained on group unions can serve any scope and any cache.

\subsection{Objectives and Exact Gains}\label{sec:gains}
Every objective is a sum over sets of a function of the page counts $c_{E,p}$. \SET{} minimizes $\Fconn$, \CLQ{} and \ARC{} maximize the pairwise objectives of Definition~\ref{def:pairwise}, and \textsc{P05} minimizes SHP's probabilistic fanout~\cite{kabiljo2017shp} at $\theta=\frac12$, $\sum_E\omega_E\sum_{p\in\pi(E)}(1-2^{-c_{E,p}})$. With integer multiplicities and a scale of $2^{12}$, its gains are exact integers for $B\le12$. Moving $v$ from page $p$ to page $p'$ changes $\Fconn$ by
\begin{equation}\label{eq:move}
\Delta(v;p\to p')=\sum_{E\ni v}\omega_E\bigl(\mathbf{1}[c_{E,p'}=0]-\mathbf{1}[c_{E,p}=1]\bigr).
\end{equation}
The first term counts sets that gain page $p'$, and the second counts sets that lose page $p$. A swap of $u$ on $p'$ and $v$ on $p$ leaves the counts of every set containing both nodes unchanged, so its exact change is
\begin{multline}\label{eq:swap}
\Delta(u\leftrightarrow v)=\Delta(v;p\to p')+\Delta(u;p'\to p)\\
+\sum_{E\ni u,v}\omega_E\bigl(\mathbf{1}[c_{E,p}=1]+\mathbf{1}[c_{E,p'}=1]\bigr).
\end{multline}
The last sum cancels the terms that the two single moves charge to such sets. The other objectives have the same structure: a move changes only the counts of the sets that contain the moved record, and a swap corrects the sets that contain both records.

\subsection{Refinement from the Incumbent}\label{sec:partition}

Algorithm~\ref{alg:hyperpage} starts from the incumbent, which transfers better than a start from scratch. Candidate pages of $v$ are the pages of its co-members within 32 positions in up to 64 of its hyperedges, plus the pages of its graph neighbors inside training hyperedges. We keep the $\bar p=16$ pages with the most such co-members. Covered nodes initiate moves, and any record on a candidate page can be a swap partner. The search is strict descent: in a fixed random node order it applies the best move or swap, by exact gain under the hard capacity, only if $\Delta F<0$. Unlike Fiduccia and Mattheyses~\cite{fiduccia1982fm}, it makes no uphill moves and no rollback, and with its candidate pages, pass limit and tolerance it certifies no local optimum. One pass takes $O(\bar pB(\mathrm{nnz}+n\deg_{\max}))$ time and $O(\mathrm{nnz}+n)$ memory, where $\mathrm{nnz}=\sum_E|E|$ and $\deg_{\max}$ is the largest number of sets that contain one node.

\begin{proposition}[Monotone refinement]\label{prop:monotone}
Algorithm~\ref{alg:hyperpage} returns $\pi$ with $F(\pi)\le F(\pi_0)$. For $F=\Fconn$, the mean reads per training scope under the scoped buffer do not exceed those of the incumbent.
\end{proposition}

\begin{proof}
Each applied operation has a negative exact change, so $F$ decreases with every operation. Corollary~\ref{cor:scope} gives mean training reads $\Fconn+1$.
\end{proof}

Proposition~\ref{prop:monotone} rules out regressions on training scopes only. Selection on validation scopes of the new run, defined in Section~\ref{sec:budget}, rules them out on the scored validation steps and promises nothing beyond them.

\subsection{Refinement and Selection for a Declared Budget}\label{sec:budget}
A declared budget fixes which records the cache will hold, and the training objective should not reward placing them. For a budget of $D$ page slots and a hot-record share $\rho$, the hybrid cache $\Ehyb{D}{\rho}$ holds the set $C$ of the $B\lfloor\rho D\rfloor$ records most frequent in training (Definition~\ref{def:budget}).

\begin{corollary}[Reads under a record cache]\label{cor:hot}
Let A1 to A4 hold for a scope $\sigma$, and let a fixed set $C$ of records be served from memory. Then $\Scoped{\sigma}$ reads exactly $\lambda_\pi(E_\sigma\setminus C)$ pages for every $\pi\in\Pi_B$ and every arrival order.
\end{corollary}

\begin{proof}
A fixed cache replaces each $e_q$ by $e_q\setminus C$ and changes nothing else, so Corollary~\ref{cor:scope}(a) applies to the sets $e_q\setminus C$, whose union is $E_\sigma\setminus C$.
\end{proof}

By Corollary~\ref{cor:hot}, a record of $C$ adds nothing to the reads of any scope, wherever it is placed. Cache-aware refinement therefore runs Algorithm~\ref{alg:hyperpage} on the residual unions $E\setminus C$ with their weights. Records of $C$ then belong to no set. They never initiate a move, and their page slots serve as free partners for swaps among uncached records. A second variant also freezes the records outside the training unions on the incumbent's pages, because the logs carry no evidence about them. We set $\rho$ to the share that the incumbent chooses on validation steps, which is known before refinement.

The misses of the page LRU of $\Ehyb{D}{\rho}$ depend on the arrival order, which the unions discard. Selection closes this gap with trace replay. For a declared $D$, every candidate layout, the incumbent included, is replayed under $\Ehyb{D}{\rho}$ in the deployment's arrival order, which the engine can log, on validation steps of the new run for every share $\rho$ of a grid. The deployment is the pair of layout and share with the fewest validation reads, with ties going to the incumbent. The incumbent with its best share is always a candidate, so the selected pair never reads more validation pages than the incumbent under the same budget.

\begin{table}[t]
\centering
\caption{Test workloads. Records are 644 bytes, so $B=12$ at 8\,KB. Overlap is $1-\sum_b|E_b|/\sum_q|e_q|$.}
\label{tab:datasets}
\footnotesize
\setlength{\tabcolsep}{1.5pt}
\begin{tabular}{@{}lrrrrrrr@{}}
\toprule
Workload & units & sub-queries & $|e_q|$ & $|E_b|$ & overlap & $L$ & recall@10 \\
\midrule
W3-1M (main) & 16,640 & 467,989 & 512.1 & 2,312 & 83.9\% & 500 & 0.946 \\
W3-21M & 16,640 & 467,989 & 818.4 & 4,475 & 80.6\% & 800 & 0.949 \\
W2 & 15,360 & 590,160 & 512.1 & 2,068 & 89.5\% & 500 & 0.950 \\
W1 & 29,256 & 963,624 & 1,211.4 & 11,830 & 70.4\% & 1,200 & 0.956 \\
W4 (proxy) & 41,355 & 1,078,138 & 513.8 & 1,858 & 86.1\% & 500 & 0.874 \\
DEEP1M & 7,174 & 199,998 & 49.1 & 1,264 & 7.6\% & 35 & 0.960 \\
\bottomrule
\end{tabular}
\end{table}

\subsection{Replay Engine}\label{sec:engine}
The engine patches stock DiskANN and keeps its record format, so every layout shares one engine. A scoped buffer is a hash map from page to an aligned page slot, emptied when the scope ends. An expanded node is a hit if the scope has read its page, a shared read if the round has already requested it, and otherwise one page read. Hits and misses stay in frontier order and no co-resident record is scored, so A1 holds, and Corollary~\ref{cor:scope} gives the reads. A relayout writes the records in the order of $\pi$ and stores a map from node to page and slot, so identifiers, PQ codes and the entry point are identical across layouts. For read-once scopes, threads take whole groups and run their logged sub-queries in order. For the hybrid cache $\Ehyb{D}{\rho}$, the hot records enter DiskANN's node cache, and every other page passes through one LRU page cache shared by all threads. A page stays pinned while a round uses it, and a request for a page in flight waits for that read. Sub-queries arrive turn by turn, and threads take them dynamically. With one thread, the engine performs the replay's cache operations in the replay's order.

\section{Experiments}\label{sec:exp}

We ask when a learned layout reads fewer pages than MARGO under an equal cache budget at both scales (\textbf{RQ1}), which design choices produce the saving (\textbf{RQ2}), and how far the layouts are from the optimum (\textbf{RQ3}).

\begin{table*}[t]
\centering
\caption{Saving in page reads over MARGO under equal cache budgets (\%, higher is better, mean of 3 draws, std below 0.5). The first row gives MARGO's reads per test step. ``On residual unions'' gives a method \HP{}'s cache-aware input. Proximity and \HEAT{} are our ports. Bold marks the best per column. The last row gives the budget, as a multiple of the declared one and a bracket on the budget grid, at which MARGO with its validation share reads as few pages as \HP{}.}
\label{tab:hybrid}
\footnotesize
\setlength{\tabcolsep}{4.2pt}
\begin{tabular}{@{}lrrrrrrrrrr@{}}
\toprule
& \multicolumn{3}{c}{W3-1M} & \multicolumn{4}{c}{W3-21M (full corpus)} & \multicolumn{3}{c}{W4-1M (layouts from W3)} \\
\cmidrule(lr){2-4}\cmidrule(lr){5-8}\cmidrule(l){9-11}
Budget & 1\% & 5\% & 25\% & 1\% & 5\% & 15\% & 25\% & 1\% & 5\% & 25\% \\
\midrule
MARGO~\cite{yue2025margo}, reads per step & 880.4k & 594.1k & 220.1k & 1,158.0k & 561.3k & 168.9k & 113.8k & 817.0k & 505.4k & 200.6k \\
\midrule
\STAR{}~\cite{wang2024starling} & 2.8 & 2.8 & 2.7 & 3.6 & 5.3 & 4.5 & 4.3 & 1.3 & 2.1 & 1.9 \\
RCM~\cite{cuthill1969bandwidth} & $-$22.6 & $-$19.3 & $-$16.8 & $-$16.1 & $-$27.1 & $-$20.0 & $-$15.7 & $-$15.1 & $-$15.7 & $-$14.1 \\
Proximity~\cite{zhao2026veloann} & $-$3.3 & $-$3.9 & $-$4.3 & $-$8.1 & $-$11.9 & $-$10.7 & $-$9.2 & 0.1 & $-$0.8 & $-$2.0 \\
\FREQ{} & $-$21.1 & $-$14.2 & $-$5.4 & $-$19.6 & $-$18.9 & $-$15.4 & $-$11.3 & $-$15.6 & $-$11.2 & 0.2 \\
\SHPG{}~\cite{kabiljo2017shp} & 0.1 & 0.0 & $-$0.1 & $-$0.2 & $-$0.3 & $-$0.1 & 0.0 & 0.0 & 0.0 & $-$0.1 \\
Mt-KaHyPar~\cite{gottesburen2024mtkahypar} & 7.4 & 1.8 & $-$4.7 & $-$5.1 & $-$10.7 & $-$14.8 & $-$10.7 & 7.5 & 4.3 & $-$1.8 \\
\quad on residual unions & 9.2 & 4.9 & 5.9 & $-$3.9 & $-$5.2 & $-$7.8 & $-$2.2 & 10.0 & 1.8 & 6.7 \\
\HEAT{}~\cite{zhong2025lsmvec} & 11.6 & 6.7 & 3.2 & 8.6 & 9.4 & $-$0.1 & $-$2.1 & 10.5 & 7.9 & 4.0 \\
\quad on residual unions & 13.4 & 10.3 & 11.2 & 9.6 & 13.0 & \textbf{5.2} & 2.6 & 12.3 & 5.2 & 10.9 \\
\midrule
\HP{} & \textbf{21.2} & \textbf{16.4} & \textbf{16.6} & \textbf{12.5} & \textbf{16.9} & \textbf{5.2} & \textbf{6.8} & \textbf{19.4} & \textbf{14.0} & \textbf{14.5} \\
\quad MARGO's budget to match ($\times$) & 2.5--3.0 & 1.4--1.6 & 1.12--1.16 & 1.5--1.75 & 1.15--1.2 & $\le$1.07 & $\le$1.08 & 2.0--2.25 & 1.4--1.6 & 1.12--1.2 \\
\bottomrule
\end{tabular}
\end{table*}

\subsection{Experiments Setup}\label{sec:exp-setup}

\subsubsection{Datasets}
Table~\ref{tab:datasets} lists the workloads. The main workload W3 comes from two public runs of the Search-R1 task~\cite{jin2025searchr1}\footnote{\label{fn:prm}\url{https://huggingface.co/datasets/wckwan/PRM-agent-rl-artifacts}} with Qwen3-8B. They search one corpus with one tool and differ in reward: an outcome reward and a process reward from an LLM verifier. Training uses steps 241 to 400 of the outcome run, in three draws by step. Validation uses steps 41 to 70 of the process run (3,836 groups), and test uses steps 71 to 200 (16,640 groups). Prompt sets align between the runs on all 400 steps, and no validation or test prompt occurs in a training step. We replay W3 on two indexes of the 21,015,324 DPR passages~\cite{karpukhin2020dpr} that the task retrieves from. W3-1M indexes a uniform sample of 1M passages, and W3-21M indexes all of them in 1,751,277 pages, 13.36\,GiB at 8\,KB. The retrieval server of these runs is not public, so we re-issue every query string of both runs through our disk-resident index and log the expanded sets, which by A1 hold under every layout. The 683\,MB index of W3-1M fits in DRAM, so at that scale the budget is a declared operating point.

Four further workloads test the direction of the result: deep-research sessions of OpenResearcher~\cite{li2026openresearcher} (W1), the outcome run alone (W2), proxy groups from OLMo-3-7B trajectories\textsuperscript{\ref{fn:prm}} (W4), and random batches of W3 sizes on DEEP1M~\cite{babenko2016deep,simhadri2022bigann}. W2 shaped the design of W3, so it does not count as confirmation. All texts are embedded with bge-small-en-v1.5~\cite{xiao2024cpack}, 384-dimensional and quantized to int8.

\subsubsection{Metrics}
The primary metric is the replayed page reads per test step under the hybrid cache $\Ehyb{D}{\rho}$ of Definition~\ref{def:budget}, kept across the 130 test steps from an empty start. Replay uses turn lockstep, which interleaves the rollouts of a step turn by turn as the official Search-R1 driver does. Each layout gets its own hot-record share $\rho$ from $\{0,0.1,0.25,0.5,0.75,0.9,1\}$: the share with the fewest validation reads after a warm-up of 10 validation steps. The W3 splits and the candidate arms other than \PFG{} and the 1M \ARCC{} were fixed before any W3 layout was trained. \PFG{} was added after the other 1M results were known, and \ARCC{} at 1M after all of them under an admission rule fixed before it was trained. \HEAT{}'s cache-aware layouts, MARGO on the uncached records and \STAR{} came after all results, the last two in Table~\ref{tab:hybrid} only. Without \PFG{}, \PFC{} and the 1M \ARCC{}, \HP{} saves 19.8\%, 14.8\% and 13.0\% at 1M. Without the three added last, it saves 3.1\% and 1.9\% at 15\% and 25\% of the full corpus. For a cache-aware layout, MARGO's validation share fixes the hot set that refinement removes, and selection then uses the same validation steps. The test steps enter neither choice. The secondary metric is the mean reads under read-once scopes $\Scoped{\sigma}$, which Corollary~\ref{cor:scope} makes exact. Recall is identical for every layout at fixed $L$, and we choose the smallest $L$ with validation recall@10 of at least 0.95, $L=500$ at 1M and $L=800$ on the full corpus. The margin of layout $A$ over MARGO is $M=1-R_A/R_{\mathrm{MARGO}}$ for total reads $R$ averaged over the three training draws. A patched DiskANN reproduces the stock results and expansion sequences of 7,850 W3 sub-queries. With one thread, it reads exactly the replayed pages of every 1M validation step under the hybrid cache. On the full corpus, near-ties of PQ distances change 44 of 66,375 validation calls, and the engine's read savings stay within 0.1 points of the replay.\looseness=-1

We test each margin with a paired bootstrap over the test steps with 2,000 resamples, also with moving blocks of 5, 10 and 26 steps. In Table~\ref{tab:hybrid}, every margin of \HP{} and 85 of the 90 margins of the compared methods differ from zero at level 0.05. Among the other rows, only \CLQG{} at 25\% of the 1M index and MARGO on uncached records at 5\% of the full corpus lose it. At 1M under the keyed group buffer, the margins of \SETG{} over MARGO and of the first three ablation steps are positive at level 0.05 under a bootstrap over prompts. The text's savings for the 1M candidate ablation, measured memory, the extra full-corpus step windows, W1, W2, smaller pages and the result cache are means of per-draw savings. The other numbers are ratios of means, and neither kind has a test.\looseness=-1

\subsubsection{Compared methods}
All layouts share the graph, record format and page size, start from MARGO's page order unless stated, and are replayed on the same buffers and caches. Residual unions drop the records that the budget's cache holds.
\begin{itemize}\setlength{\itemsep}{0.5pt}
\item MARGO~\cite{yue2025margo}: monotonic-path arc weights, no logs, from its authors' code\footnote{\url{https://github.com/CodenameYZY/MARGO}}. It also runs in its own page-search engine, which breaks A1.
\item \STAR{}~\cite{wang2024starling}: Starling's in-page arcs, refined from the identifier order.
\item RCM~\cite{cuthill1969bandwidth} and Proximity: the reverse Cuthill-McKee order of the graph, and recursive balanced 2-means on the vectors in the spirit of VeloANN~\cite{zhao2026veloann}, both without logs.
\item \FREQ{}: records packed by training frequency, our baseline.
\item \HEAT{}: our port of LSM-VEC's heatmap~\cite{zhong2025lsmvec}, logged arc traversals plus one per arc. LSM-VEC specifies neither its weights nor its window and releases no code, so we refine a layout under its arc weights with our optimizer, in the weighting that reads fewer validation pages.
\item Mt-KaHyPar~\cite{gottesburen2024mtkahypar}: connectivity at exact page capacity, reported on group unions and on residual unions. On group unions it partitions in two levels as in Bandana~\cite{eisenman2019bandana} at 1M. At 21M it refines MARGO's layout with its V-cycles within bands of 43 consecutive pages, because one hypergraph of all pages exceeds 90\,GB. On residual unions it takes its variant, one run or two levels, with fewer validation reads.
\item \SHPG{}: SHP's paired moves~\cite{kabiljo2017shp} on the \textsc{P05} objective, the move rule only.
\item Algorithm~\ref{alg:hyperpage}: \ARCG{}, \CLQG{}, \SETG{} and \PFG{} under \ARC{}, \CLQ{}, \SET{} and \textsc{P05} on group unions (cache-blind), \ARCQ{} and \SETQ{} on per-query sets, \PFC{} and \ARCC{} on the residual unions of one budget (cache-aware), and \ARCCF{}, which is \ARCC{} with the records outside the training unions frozen.
\item \HP{}: the pair of layout and share with the fewest validation reads among MARGO, \STAR{}, Algorithm~\ref{alg:hyperpage}'s layouts for the budget, and \HEAT{}'s objective and MARGO's partitioner on the uncached records.
\end{itemize}
The layouts of Algorithm~\ref{alg:hyperpage} share the start, candidate pages, passes and tolerance, so only the objective and the hyperedge family differ. The first cache-aware layout of each scale uses the objective that leads the cache-blind layouts on its validation steps, and \ARCC{} is also a candidate at 1M.

\subsubsection{Implementation details}
At both scales, the fixed Vamana graph is built with MARGO's DiskANN-derived builder, with $R=64$, build list size 125 and 32-byte PQ codes. Traversals are logged by stock DiskANN\footnote{\url{https://github.com/microsoft/DiskANN}} with $W=4$. Each learned layout is built once per training draw, with at most 8 passes and $\epsilon=10^{-4}$, and tables report mean $\pm$ std over the 3 draws. On one thread, eight passes of a cache-blind layout take 1,991 to 13,830\,s, and cache-aware refinement 317 to 5,943\,s, because the residual unions keep only 8\% to 73\% of the incidences. Timing uses one consumer NVMe SSD per host with direct I/O and 12 search threads pinned to their own cores. Both arms of a timed pair hold the same hot records and LRU pages, except at 1\% of 1M, where each arm uses its own share. DiskANN's node cache spends 1.25 to 1.72 times the declared share on the hot records of every layout. Our code is publicly available: \url{https://github.com/zhangshea/HyperPage}

\subsection{Overall Comparison}\label{sec:exp-results}

\paragraph*{Overall}
\HP{} reads fewer pages than MARGO in every column of Table~\ref{tab:hybrid}, and no compared layout reads fewer. It saves 16.4\% to 21.2\% at 1M and 12.5\% and 16.9\% at 1\% and 5\% of the full corpus. At 1\% and 5\%, every layout of Algorithm~\ref{alg:hyperpage} in Table~\ref{tab:components} also beats MARGO at both scales, so the logs carry most of the gain when the cache is small. At 15\% and 25\% of the full corpus, the cache serves most test accesses, and the saving falls to 5.2\% and 6.8\%.

\paragraph*{A layout substitutes for memory}
A better layout reaches MARGO's reads with a smaller cache (last row of Table~\ref{tab:hybrid}). MARGO needs 2.5 to 3.0 times the declared budget to read as few pages as \HP{} at 1\% of the 1M index, and 1.5 to 1.75 times at 1\% of the full corpus. The factor shrinks as the budget grows. At 1\% of the 21M index, the extra 68 to 103\,MiB are 5\% to 8\% of the engine's peak resident memory, so at this scale the layout saves page reads more than memory.{} Charged the measured memory of DiskANN's node cache instead of the declared share, \HP{} still saves 18.20\% to 29.02\% at 1M on every draw.

\begin{figure}[t]
\centering
\includegraphics[width=\columnwidth]{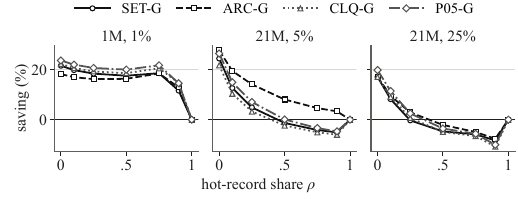}
\caption{Page-read saving over MARGO on W3 test when every layout uses the same hot-record share $\rho$, with no selection (higher is better). Panels give the index and the budget.}
\label{fig:rho}
\end{figure}

\paragraph*{The budget must enter refinement}
Logged layouts that ignore the cache lose to MARGO at the large budgets of the full corpus, because they spend their gain on cached records. Under a pure page LRU, our four cache-blind objectives beat MARGO at all seven budgets, by 14.22\% to 40.02\% ($\rho=0$ in Fig.~\ref{fig:rho}, which shows three budgets). At 25\% of the full corpus, they read more than MARGO for every share from 0.5 to 0.9. At 15\% and 25\%, they read 3.5\% to 11.5\% more at their validation shares. These shares cache 11.2\% and 22.5\% of the records, which serve 74.8\% and 84.7\% of the test accesses and carry 79\% and 92\% of the training incidences. By Corollary~\ref{cor:hot}, these records add nothing to the reads of any scope, and leaving them out of refinement reverses the loss (Section~\ref{sec:exp-ablation}). A cache-aware layout is tied to its budget: at the other large budget it reads 0.8\% to 3.0\% more than MARGO on validation steps, so \HP{} keeps the incumbent among the candidates. A cache-aware layout also needs a record cache. If the share holds the most accessed pages instead, \PFC{}'s saving falls to 8.43\%, $-3.33\%$ and 1.56\% at 1M.

\paragraph*{Why the baselines fall short}
Orders without logs gain little, and no log-aware baseline reads fewer pages than \HP{} in any column. \STAR{} saves at most 2.8\% at 1M and 3.6\% to 5.3\% on the full corpus, second only to \HP{} at 25\%. Proximity and RCM read more pages than MARGO in all but one cell. \HEAT{}, deployed cache-aware on every draw, saves 5.4 to 7.8 points less than \HP{} on W3-1M, and 3.0 and 4.0 points less at 1\% and 5\% of the full corpus. Our residual unions raise both Mt-KaHyPar and \HEAT{} in 9 of 10 columns, the exception being W4 at 5\%. Mt-KaHyPar on the same residual unions saves 10.7 to 12.1 points less on W3-1M. On the full corpus it reads 2.2\% to 7.8\% more than MARGO, although it lowers the connectivity of the training unions by 18\% to 26\%. \PFC{}'s layouts also have 11.7\% to 12.0\% lower connectivity of the residual training unions than the two-level partitioner's, its own objective, so the gap lies in optimization at exact capacity. Training connectivity is thus an unreliable guide to held-out reads. Under one shared cache, residual training connectivity ranks the validation reads of each W3 column's 18 layouts with Spearman's $r_s$ of 0.06 to 0.92. Validation replay ranks test reads at 0.92 to 1.00, so \HP{} selects by replay.

\begin{figure}[t]
\centering
\includegraphics[width=\columnwidth]{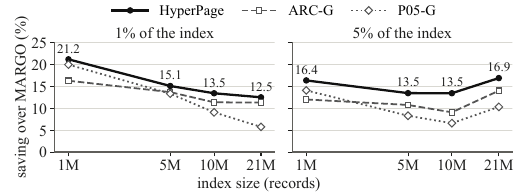}
\caption{Saving in page reads over MARGO on W3 test as the index grows (\%, higher is better, 3 draws). Every layout uses its own validation share. \HP{} picks among MARGO, \ARCG{}, \PFG{}, \ARCC{} and \PFC{}, which reproduces its cells of Table~\ref{tab:hybrid} at 1M and 21M.}
\label{fig:scale}
\end{figure}

\paragraph*{Scale}
The saving persists as the index grows (Fig.~\ref{fig:scale}, with uniform 5M and 10M samples of the corpus and the same queries). At every size from 1M to 21M, \HP{} reads 12.5\% to 21.2\% fewer pages than MARGO at 1\% and 5\% and saves more than either cache-blind objective. At 1\%, its saving falls by 6.1 points from 1M to 5M and by 2.6 more to the full corpus, consistent with the larger share of accesses that 1\% of a larger index serves (Section~\ref{sec:intro}). The leading objective also changes, from \PFG{} at 1M to \ARCG{} on the full corpus, where the training unions cover 33\% of the records instead of 86\%.

\paragraph*{Stability over the run}
A layout learned once keeps its advantage across the evaluation run. From step 1 to step 240, the savings of the frozen 1M deployments fall by only 2.0 to 2.6 points. On full-corpus steps 1 to 40 and 201 to 240, which no selection used, the frozen full-corpus deployments read 4.7\% to 22.1\% and 4.9\% to 12.5\% fewer pages than MARGO. Each also beats every other candidate at its validation share on every draw. We fix the graph, so inserts and deletes stay outside this study.

\paragraph*{Search time}\label{sec:exp-time}
\HP{}'s fewer page reads cut the search time in the hybrid-cache engine at every budget we time (Table~\ref{tab:time}). The median paired saving is 9.60\% to 13.87\% per step at 1M and 2.59\% to 10.38\% on the full corpus. The table times MARGO and \HP{}'s timed layout of each budget over the 130 test steps, in alternating pairs. A pair is void when a check fixed before the run flags host load, CPU affinity or the read speed of the two files. Two pairs were void, one at 25\% of 1M and one at 1\% of the full corpus. At every budget, \HP{} is faster in every valid pair, and each budget with 10 pairs gives a sign-test $p$ of 0.002. The time saving is smaller than the read saving because the engine reads the pages of one beam round in parallel, so time follows the rounds that touch the disk. Median sub-query latency falls at every budget, by 2.2\% to 16.1\%, and p99 falls by 2.1\% to 6.3\%. At 1\% and 5\% of 1M under the lockstep shares, 12 concurrent sub-queries share a 417-page LRU, and the engine reads 31\% to 38\% more than the replay. Replayed in the arrival order that the engine logs with MARGO's layout, MARGO's reads match the engine's within 0.001\%. The timed 1M layouts at 1\% and 5\% are selected under this order, which approximates each candidate's own order. They were selected before \STAR{}, \ARCCF{} and the \HEAT{} and MARGO layouts on uncached records joined the candidates. On the full corpus, where the LRU holds thousands of pages, the engine reads 0.64\% and 0.38\% fewer pages than the replay at 1\% and 5\%. Both arms of every pair reach the same peak resident memory within 0.4\%, PQ codes and engine buffers included. The public logs carry no step timing, so we cannot report retrieval's share of a training step, and these times measure the retrieval tier alone. Payback charges the refinement in core-seconds and the index rewrite. With 400 steps per run, a layout refined once pays for itself within 0.8 to 3.1 later runs in core-seconds at 1M and at 1\% and 5\% of the full corpus. In 12-thread wall time it needs 9.7 to 37.5 runs there. At 15\% and 25\% of the full corpus, where it saves 2.59\% and 3.48\% of the time, it needs 45.9 and 6.6 runs in core-seconds and 551 and 79 runs of wall time.

\begin{table}[t]
\centering
\caption{Search time per W3 test step in the hybrid-cache engine (12 threads, one host). Time saved is the median of the paired savings. The 1M rows at 1\% and 5\% time layouts selected in the engine's arrival order, not the deployments of Table~\ref{tab:hybrid}. Rows 21M 1\% and 5\% come from rounds of five arms, so their two arms are not adjacent.}
\label{tab:time}
\footnotesize
\setlength{\tabcolsep}{3pt}
\begin{tabular}{@{}llccccc@{}}
\toprule
& & & \multicolumn{2}{c}{s per step} & reads & time \\
\cmidrule(lr){4-5}
Budget & layout & pairs & MARGO & arm & saved\% & saved\% \\
\midrule
1M, 1\% & \CLQG{} & 10 & 5.74 & 5.00 & 18.82 & 12.85 \\
1M, 5\% & \PFC{} & 10 & 3.73 & 3.21 & 20.38 & 13.87 \\
1M, 25\% & \ARCC{} & 10 & 1.79 & 1.62 & 16.60 & 9.60 \\
\midrule
21M, 1\% & \ARCC{} & 3 & 6.58 & 6.06 & 12.65 & 7.80 \\
21M, 5\% & \ARCC{} & 3 & 3.98 & 3.57 & 17.04 & 10.38 \\
21M, 15\% & \HEAT{}, res. & 10 & 2.11 & 2.06 & 5.28 & 2.59 \\
21M, 25\% & MARGO, uncached & 10 & 1.77 & 1.71 & 6.72 & 3.48 \\
\bottomrule
\end{tabular}
\end{table}

\paragraph*{MARGO's page-search engine}\label{sec:exp-ps}
The hybrid engine with \HP{}'s layout is the fastest configuration we time, and \HP{} also speeds up MARGO's own engine at 1M (Table~\ref{tab:ps}). MARGO's engine scores every record of a fetched page, so its walks depend on the layout and fall outside A1. We port it to 8\,KB pages with a record cache of the whole budget ranked by training frequency, which beats its own cache selection, and refine \HP{} on the records that this cache leaves uncached. At 1M, \HP{}'s page-search layouts read 11.8\%, 8.3\% and 7.5\% fewer pages than MARGO's on every draw, at test recall within 0.3 points. With either layout, page search takes 1.06 to 2.19 times as long as the hybrid engine and holds 20\% to 33\% more memory. At 1M, \HP{}'s layouts take 9.8\% to 14.0\% less time than MARGO's in the hybrid engine and 1.9\% to 10.6\% less in page search, in every round. On the full corpus, its page-search layout takes 8.2\% less time at 1\%. At 5\% it needs $L=700$ instead of 600 to reach the recall target, so it reads 11.4\% more pages and takes 13.6\% more time. In the hybrid engine, \HP{} takes 1.44\% and 2.62\% less time than \HEAT{} on residual unions, the fifth arm on the full corpus. At 5\% with 64 to 256 concurrent sub-queries, it serves about twice the sub-queries per second of page search with MARGO's layout, at lower p99 latency. MARGO's hybrid runs slowed within each run at this load, so we leave them out.

\begin{table}[t]
\centering
\caption{MARGO's page-search engine against the hybrid cache (12 threads, the host of Table~\ref{tab:time}): page reads per test call / seconds per test step, medians of 3, 4 and 2 valid rounds at 1M and 3 at 21M. Page search takes the smallest $L$ with validation recall@10 of 0.95. Test recall@10 is 0.946 to 0.954. Its 1M hybrid arms at 1\% and 5\% are those of Table~\ref{tab:time}. At 25\% it times \PFC{} on older copies of the files, where MARGO takes 1.85\,s per step instead of 1.79\,s.}
\label{tab:ps}
\footnotesize
\setlength{\tabcolsep}{2.5pt}
\begin{tabular}{@{}lcccc@{}}
\toprule
& \multicolumn{2}{c}{Page search} & \multicolumn{2}{c}{Hybrid cache} \\
\cmidrule(lr){2-3}\cmidrule(lr){4-5}
Budget & MARGO & \HP{} & MARGO & \HP{} \\
\midrule
1M, 1\% & 200.3 / 6.08 & 176.5 / 5.44 & 331.3 / 5.74 & 269.0 / 5.00 \\
1M, 5\% & 149.6 / 4.83 & 137.0 / 4.53 & 175.0 / 3.73 & 139.4 / 3.21 \\
1M, 25\% & 78.4 / 3.08 & 72.4 / 3.02 & 60.7 / 1.85 & 51.1 / 1.67 \\
21M, 1\% & 296.9 / 9.66 & 268.9 / 8.88 & 319.7 / 6.58 & 279.3 / 6.06 \\
21M, 5\% & 187.2 / 6.89 & 208.4 / 7.82 & 155.4 / 3.98 & 128.9 / 3.57 \\
\bottomrule
\end{tabular}
\end{table}

\paragraph*{Transfer to another agent}\label{sec:exp-transfer}
Layouts learned from one agent's run also serve another agent, but the cache split must be chosen anew. The W4 columns of Table~\ref{tab:hybrid} replay the held-out traversals of an OLMo-3-7B agent at eight checkpoints on the same 1M index (test recall@10 0.874). Every layout and hot-record ranking comes from the W3 training run, and only the share and the deployed layout are chosen on W4 validation steps. \HP{} reads 19.4\%, 14.0\% and 14.5\% fewer pages than MARGO at 1\%, 5\% and 25\% on every draw, and its saving moves by at most 1.21 points from the first checkpoint to the last. At 5\%, MARGO's share moves from 0.90 on W3 to 0.75 on W4, so selection deploys \PFG{} instead of the layouts refined for W3's hot set.

\paragraph*{Read-once scopes}\label{sec:exp-scope}

Under read-once buffers, every objective of \HP{} beats MARGO at both scales, and only Mt-KaHyPar comes close, at 1M (Table~\ref{tab:scope}). Under the keyed group buffer, our objectives read 14.99\% to 20.81\% fewer pages than MARGO at 1M and 9.64\% to 13.65\% fewer on the full corpus. Mt-KaHyPar saves 12.44\% at 1M but 0.63\% on the full corpus. A keyed buffer needs memory, however: the 128 live groups of a step hold 144,703 pages at 1M, more than the 83,334-page index, and 2.54 to 2.95\,GiB on the full corpus.

\subsection{Ablation Study}\label{sec:exp-ablation}

Both steps of \HP{} matter, and the budget decides which one carries the saving (Table~\ref{tab:components}). Learning from logs alone, the better cache-blind layout saves 8.2\% to 20.0\% at 1M and at 1\% and 5\% of the full corpus. Refining \ARC{} cache-aware adds 1.2 to 11.9 points in every W3 column and turns the losses at 15\% and 25\% of the full corpus into savings. Even without selection, this fixed recipe, \ARCC{}, beats MARGO in every W3 column, by 1.2\% to 18.1\%. At these two budgets, most of the gain comes from leaving cached records out rather than from the logged objective. MARGO on uncached records partitions the graph induced by those records with MARGO's code and fills the free slots with the cached ones. It saves 4.6\% and 6.8\% there, while the best logged objective adds 0.6 points at 15\% and loses 4.2 at 25\%. Elsewhere it saves at most 5.9\%, and at 5\% of the full corpus it reads 0.9\% more than MARGO, a margin not significant with moving blocks. No layout is best in every column, so \HP{} selects on validation, and in every column its pick is also the best of its candidates on test.

\begin{table}[t]
\centering
\caption{Reads under read-once scopes on W3 test: mean reads per group and the saving $\Delta$ over MARGO (\%) per group and per tool call. Bold marks the best and underline the second best per column.}
\label{tab:scope}
\footnotesize
\setlength{\tabcolsep}{1.8pt}
\begin{tabular}{@{}lcccccc@{}}
\toprule
& \multicolumn{3}{c}{W3-1M} & \multicolumn{3}{c}{W3-21M} \\
\cmidrule(lr){2-4}\cmidrule(l){5-7}
Layout & group$\downarrow$ & $\Delta$ & call $\Delta$ & group$\downarrow$ & $\Delta$ & call $\Delta$ \\
\midrule
\STAR{} & 1,408.1 & $-$0.22 & 1.32 & 2,944.6 & 2.40 & 3.01 \\
MARGO & 1,405.0 & 0.00 & 0.00 & 3,017.0 & 0.00 & 0.00 \\
\FREQ{} & 1,622.5 & $-$15.47 & $-$17.76 & 3,439.1 & $-$13.99 & $-$17.31 \\
Mt-KaHyPar & 1,230.2 & 12.44 & 10.58 & 2,998.1 & 0.63 & $-$0.46 \\
\SHPG{} & 1,404.0 & 0.08 & 0.09 & 3,019.2 & $-$0.07 & $-$0.09 \\
\midrule
\ARCG{} & 1,194.4 & 14.99 & 17.08 & \textbf{2,605.1} & \textbf{13.65} & \textbf{15.02} \\
\CLQG{} & 1,147.8 & 18.31 & \underline{20.53} & 2,726.1 & 9.64 & 11.20 \\
\SETG{} & \underline{1,130.5} & \underline{19.54} & 19.76 & 2,659.8 & 11.84 & 11.70 \\
\PFG{} & \textbf{1,112.7} & \textbf{20.81} & \textbf{21.77} & \underline{2,644.5} & \underline{12.35} & \underline{13.07} \\
\bottomrule
\end{tabular}
\end{table}

Within the cache-blind step, logging at all gives most of the saving, and each refinement of the objective adds a little. Under read-once scopes at 1M, an arc-restricted objective on logged group unions saves 14.99\% against MARGO. Dropping the arc restriction adds 3.90\%, as the off-arc share of co-located pairs rises from 0.404 to 0.787. The set form and \textsc{P05} add 1.51\% and 1.58\% more with the same start, candidates and passes, the first consistent with Propositions~\ref{prop:bonus} and~\ref{prop:ratio}. The move rule matters too: on the \textsc{P05} objective from the same start, SHP's move rule saves 0.08\% where \PFG{} saves 20.81\%. Of the gain that SHP predicts on a round's snapshot, 78\% to 80\% does not materialize.

The logs mostly supply the query distribution and the co-access inside sub-queries. Refining on sub-queries shuffled across the groups of a step keeps 16.28\% of the 19.54\% at group scope. {} An exact-string result cache of at most 1/64 of the budget complements the layout: it saves MARGO 16.82\% to 25.17\% of its reads and leaves \HP{} 12.14\% to 20.59\% ahead.

\paragraph*{Other workloads and page size}\label{sec:exp-other}

The saving holds on other agent workloads and smaller pages, and it depends on how much the queries of a batch overlap. On W1 and W2 under the hybrid cache, the candidates are MARGO, the four cache-blind layouts and \PFC{}, and MARGO's share is chosen per draw. \HP{} then reads 16.71\% to 24.24\% fewer test pages than MARGO on W1 and 20.50\% to 25.38\% fewer on W2. On DEEP1M, whose random batches overlap by 7.6\%, the cache-blind layouts save only 0.33\% to 4.74\%, against 18.62\% for \ARCG{} at 1\% on W1. With 2 to 6 records per page and candidates refined anew, \HP{} still reads less than MARGO on W3-1M on every draw. At 1\% and 5\%, its saving falls with $B$, since each record has at most $B-1$ page partners: it saves 13.96\% and 12.00\% with 6 records per page and 4.49\% and 5.56\% with 2.

\subsection{Optimization Audit}\label{sec:exp-audit}
The learned layouts read at most 2.05 to 2.20 times the optimum under read-once group scopes, against 2.59 times for MARGO, but the bound leaves the true gap open. On the cold read-once group scopes of Table~\ref{tab:scope}, Proposition~\ref{prop:lb} with 67 record classes gives 543.2 reads per W3-1M test group, 2.81 times the capacity bound $L_{\mathrm{cap}}$. The best layout we found, refined on the test groups themselves and so only an in-sample reference, reads 974.4, and longer runs changed it by under 0.5\%. Hence $543.2\le R^\star\le974.4$, which gives the ratios above. We also solve 71 smaller W3 instances to proven optimality, with 12 to 48 records and $B$ from 2 to 12. On them, the bound with one class per record reaches a median of 96\% of the optimum at $B=3$ and 81\% to 90\% at $B=12$. The bound is thus nearly tight at small $B$, and its looseness at $B=12$ leaves open how much a better layout could still save. Under the shared cache of Table~\ref{tab:hybrid}, the window bound of Proposition~\ref{prop:window} lies 11 to 48 times below \HP{}'s reads, so the optimum of the primary metric remains open.

\begin{table}[t]
\centering
\caption{Where \HP{}'s saving comes from: saving over MARGO (\%) of \HP{}'s candidates. Bold marks \HP{}'s pick on validation.}
\label{tab:components}
\scriptsize
\setlength{\tabcolsep}{2.6pt}
\begin{tabular}{@{}lrrrrrrr@{}}
\toprule
& \multicolumn{3}{c}{W3-1M} & \multicolumn{4}{c}{W3-21M} \\
\cmidrule(lr){2-4}\cmidrule(l){5-8}
Layout & 1\% & 5\% & 25\% & 1\% & 5\% & 15\% & 25\% \\
\midrule
\multicolumn{8}{@{}l}{\textit{Logs, cache-blind}} \\
\ARCG{} & 16.3 & 12.1 & 4.7 & 11.3 & 14.1 & $-$3.5 & $-$7.6 \\
\PFG{} & 20.0 & 14.1 & 8.2 & 5.8 & 10.3 & $-$10.0 & $-$10.0 \\
\multicolumn{8}{@{}l}{\textit{Plus cache-aware refinement}} \\
\ARCC{} & 18.1 & 15.6 & \textbf{16.6} & \textbf{12.5} & \textbf{16.9} & 3.1 & 1.2 \\
\PFC{} & \textbf{21.2} & \textbf{16.4} & 15.8 & 7.3 & 10.1 & $-$5.7 & $-$3.5 \\
\ARCCF{} & 17.0 & 14.6 & 15.2 & 8.1 & 9.7 & 2.4 & 1.9 \\
\HEAT{}'s objective & 13.4 & 10.3 & 11.2 & 9.6 & 13.0 & \textbf{5.2} & 2.6 \\
\multicolumn{8}{@{}l}{\textit{MARGO's objective, cache-aware}} \\
MARGO on uncached records & 0.3 & 1.3 & 5.9 & 0.3 & $-$0.9 & 4.6 & \textbf{6.8} \\
\bottomrule
\end{tabular}
\end{table}

\section{Conclusion}\label{sec:conclusion}
The logs of one finished agent run price every page layout of a fixed disk-resident graph index, and cached records add no page reads. Choosing a layout for a declared budget thus becomes a measurable optimization problem. \HP{} builds layouts on the uncached part of the logged traversals and selects one by replay. On the main workload it reads up to 21.2\% fewer pages than MARGO, most when the cache is small. Once the cache serves most accesses, the budget must enter refinement. A Lagrangian bound brackets the remaining gap to the optimum. Index updates, larger corpora and the optimum under a shared cache remain open.

\bibliographystyle{IEEEtran}
\bibliography{refs}

\end{document}